\documentclass[aps,prd,preprint,amsmath,amssymb,superscriptaddress,floatfix]{revtex4-2}

\usepackage{graphicx}

\makeatletter
\@addtoreset{equation}{section}
\makeatother
\renewcommand{\theequation}{\arabic{section}.\arabic{equation}}
\usepackage{subcaption}

\usepackage{hyperref}
\hypersetup{colorlinks=true, linkcolor=blue, citecolor=magenta, urlcolor=blue}

\usepackage{ragged2e}

\begin{document}

	\title{Holographic Pseudo Entropy from the Rindler Method in AdS and Flat Space}
	
	\author{Mehdi Hakami Shalamzari}
	\email[Corresponding author: ]{mehdi.hakami@gmail.com, m.hakami@scu.ac.ir}
	\affiliation{Physics Department, Faculty of Science, Shahid Chamran University of Ahvaz, Ahvaz 61357-43135, Iran}
	
	\author{Zahra Paygozar}
	\email{zahrapaygozar79@gmail.com}
	\affiliation{Physics Department, Faculty of Science, Shahid Chamran University of Ahvaz, Ahvaz 61357-43135, Iran}
	
	\author{Mahdi Kord Zangeneh}
	\email{mkzangeneh@scu.ac.ir}
	\affiliation{Physics Department, Faculty of Science, Shahid Chamran University of Ahvaz, Ahvaz 61357-43135, Iran}

	\begin{abstract}
		Via the Rindler method, we study holographic timelike entanglement entropy (TEE) for a general timelike interval in AdS$_3$/CFT$_2$ and flat$_3$/CCFT$_2$. The TEE is the pseudo entropy of a timelike interval. In the bulk, the imaginary part of the TEE comes from a timelike geodesic.  After a Rindler transformation to a rotating BTZ black hole, the two spacelike extremal curves lie on the inner and outer horizons, and exchanging $l_u\leftrightarrow \ell l_\phi$ takes one to the other. We clarify the geometric role of the timelike extremal curve in the Rindler frame. We show that it is the line $\tilde U=\tilde V=0$ stretched between the two horizons, starting on the inner horizon, touching the outer horizon at its midpoint, and returning to the inner one. No such curve exists outside the region between the horizons. Its length, $i\pi\ell$, is independent of the interval, and it alone produces the imaginary part $i\pi c/6$. We follow the same procedure in flat space holography, where the spacelike curve sits on the cosmological horizon and the timelike curve lies between the cosmological horizon and timelike infinity. Evaluating the holographic TEE for a timelike interval in CCFT$_2$ on the plane, we find that, in contrast with the cylinder, the real and imaginary parts are equal. 
	\end{abstract}

	\maketitle
	\thispagestyle{empty}
	\newpage

\tableofcontents

	\section{Introduction}
	
The Ryu--Takayanagi (RT) formula \cite{RT2006,RT2006b} equates the entanglement entropy of a boundary region with the area of an extremal bulk surface in Planck units, and the covariant version of Hubeny, Rangamani and Takayanagi \cite{HRT} takes care of time dependence. Few entries in the AdS/CFT dictionary \cite{Maldacena1997,GKP,Witten1998} are as plainly geometric. The formula also underlies the idea that spacetime emerges from boundary entanglement \cite{VanRaamsdonk2010,VanRaamsdonk2016}. A result this central should be derived, not just checked against examples.

Derivations do exist. The RT prescription was obtained from the bulk replica geometry \cite{Lewkowycz2013}, later extended to the covariant (HRT) case \cite{DLR2016}. For spherical regions there is a shorter route \cite{Casini2011}, and it is the one we follow. A conformal map takes the causal development of a ball to
$\mathbb{R}\times H^{d-1}$. There the vacuum is a thermal state at
$T=1/2\pi R$, so the entanglement entropy becomes a thermal entropy. In
the bulk, the same change of variables re-slices pure AdS as a black
hole with a hyperbolic horizon, and the Bekenstein--Hawking entropy of
that horizon is the area of the RT surface. The modular Hamiltonian of
the ball (interval) is local and generates a geometric flow, namely the
Rindler boost transported to the interval by the conformal map, and this
flow has a bulk extension, a Killing vector, whose fixed points form the
extremal surface. Since the conformal map is built from the Rindler
wedge, the procedure is known as the Rindler method. More recently, the RT formula has also been derived directly from
CFT$_2$ data using the large-$c$ ensemble, with the dual geometry
emerging from the CFT~\cite{Bao2504,Geng2505}.

Nothing in this construction is tied to AdS. It has been used for warped CFTs and their duals \cite{CHI,SWX}. Flat space is a closer relative for our purposes. Three-dimensional asymptotically flat gravity has the BMS$_3$ group as its asymptotic symmetry \cite{BT2010}, and flat holography proposes a dual field theory with that symmetry, a BMS-invariant theory that we call CCFT$_2$ \cite{Bagchi2010,BagchiFareghbal2012}. Entanglement entropy in such theories was computed first with twist fields \cite{Bagchi2015}, then with Wilson lines \cite{BasuRiegler}. A derivation on the metric side, again through the Rindler method, was given in \cite{Jiang_2017}, where it was found that the RT picture has to be modified. The extremal spacelike geodesic does not end on the boundary interval. Two null geodesics, the orbits of the interval endpoints under the bulk modular flow, connect it to null infinity, and only the spacelike segment contributes to the length. The Rindler map sends this geodesic to the horizon of a flat-space cosmology (FSC), and the result agrees with the flat limit of the AdS$_3$ calculation. Flat-space cosmologies are time-dependent solutions of three-dimensional flat gravity with a cosmological horizon \cite{Cornalba_2002}, and their holographic description was studied in \cite{Bagchi_2013}.

Another generalization leaves the background alone and changes the region. Continue the entropy of a spacelike interval in a two-dimensional CFT to a timelike one and you get
\begin{equation}
S^{(T)} = \frac{c}{3}\log\frac{\sqrt{\Delta t^{2}-\Delta x^{2}}}{\epsilon} + \frac{i\pi c}{6},
\end{equation}
a complex number. This timelike entanglement entropy (TEE) was defined through a Wick rotation and the replica trick, and checked numerically \cite{Doi2022,Doi_2023}; an independent treatment appeared in \cite{Li2022}. It is best read as a pseudo entropy \cite{Nakata2021}, the von Neumann entropy of a transition matrix, which explains why it is not real. Concretely, one takes two different pure states, traces out part of the system from the operator that maps one onto the other, and divides by their overlap. The transition matrix that comes out is not Hermitian, so its eigenvalues are complex in general, and so is the entropy \cite{Nakata2021}. If the two states coincide, one is back to ordinary entanglement entropy. The timelike case fits this pattern. The reduced density matrix of a timelike interval is not Hermitian, and a thermofield-double-like construction with two different states, one evolved with the opposite sign of the time parameter, reproduces it as a transition matrix \cite{Doi_2023}. The same kind of complex entropy shows up in dS/CFT, and a double Wick rotation relates it to the timelike entropy of AdS/CFT \cite{Doi2022,Doi_2023}. In Euclidean time-dependent asymptotically AdS geometries, the pseudo entropy is given by the same minimal-surface formula as ordinary entanglement entropy \cite{Nakata2021}.

Since these original works \cite{Doi2022,Doi_2023}, timelike
entanglement entropy and pseudo entropy have been explored in many
directions, including its geometric interpretation and temporal
entanglement from holographic entanglement entropy
\cite{Heller2408,Heller2507}, the interpolation between spacelike and
timelike entanglement \cite{Roy2507}, higher dimensions
\cite{Roy2508}, gravitational anomalies and higher curvature gravity
\cite{Anom2504,HC2509}, the entanglement phase transition in pseudo
entropy \cite{Kanda2311}, the entanglement first law \cite{Sun2511},
signatures of ergodicity and chaos \cite{Chaos2601}, the inner horizon
\cite{Wen2025}, and the flat-space limit and Carrollian pseudo entropy
\cite{Fareghbal:2024lqa,Fareghbal_2026,shalamzari2026modularflowholographictimelike}. Pseudo entropy has also been studied in free gauge theory
\cite{Mukherjee2205}, and timelike entanglement entropy has been
considered in the Lifshitz background \cite{Jena2410}; for
Carrollian theories, boosted intervals have been used in the study of
symmetry-resolved entanglement \cite{Banerjee2404}. 

Holographically, no single spacelike geodesic connects the endpoints of a timelike interval. One needs two spacelike geodesics running from the endpoints of the boundary interval and a timelike geodesic joining them. The spacelike pieces give the real part. The timelike piece has length $i\pi\ell$, hence the imaginary part $i\pi c/6$, and it has been linked to an emergent time coordinate \cite{Doi2022,Doi_2023}. The construction has since been pushed behind horizons, where TEE reaches regions that spacelike extremal surfaces typically do not \cite{LiYang2026}, and it has been related to the structure of the inner horizon \cite{Wen2025}. Pseudo entropy has also been used to diagnose quantum chaos and to tell quantum phases apart (see \cite{He2024} and references therein). Its physical meaning for timelike intervals is still not well understood \cite{He2024}. And complex entropies are not tied to timelike intervals alone. In a rotating BTZ background, for a particular choice of metric, they appear even for spacelike ones \cite{Doi_2023}.

A recent work asked whether the Rindler method can reproduce all of this \cite{He2024}. Its answer is partial. After a Rindler transformation with the sign of $l_v$ flipped, the TEE equals a thermal entropy plus $i\pi c/6$, and the horizon of the transformed black hole maps back to the two spacelike geodesics. A black hole entropy is real, so a horizon area cannot supply the imaginary part, and the constant is added by hand. 	
Several questions are left open. Is there a bulk flow whose zero set is
the timelike curve? Can its length, and hence the imaginary part, be
computed directly in the Rindler frame, where the spacelike curves are
horizons? And does any of it survive in flat holography?

A Rindler transformation turns the AdS$_3$ flows into Killing vectors of a rotating BTZ black hole. This is the black hole of three-dimensional gravity with a negative cosmological constant, and it has an outer and an inner horizon \cite{Ba_ados_1992,Ba_ados_1993}. The Rindler picture of the extremal curves has a longer history in
flat-space holography. The flat-space limit of the Ryu--Takayanagi
curves is not well defined in general. In \cite{Fareghbal:2020dtq} a new
vector field was introduced in AdS$_3$, vanishing on a curve whose
flat limit is well defined and which is orthogonal to the bulk modular
flow of the original interval. After a Rindler transformation the bulk
modular flow becomes the Killing vector normal to the outer horizon of
BTZ, while the new vector becomes the Killing vector normal to the
inner horizon. The two flows are exchanged by interchanging the
interval parameters $l_u$ and $\ell l_\phi$, and the flat limit of the
second one reproduces the holographic entanglement entropy of
BMS-invariant field theories. It was later shown that these curves are
the spacelike branches of the extremal curves of pseudo entropy for a
timelike interval, which gives them their role in the timelike
problem \cite{Fareghbal:2024lqa}. The flow $\xi_{st}$ used below is this
second vector.

The inner horizon was brought to the center of the timelike problem in
\cite{Wen2025}. There the preimage of the inner horizon is identified
with the spacelike geodesics of the timelike entanglement entropy, and
the corresponding modular momentum flow, its fixed-point structure, and
the associated modular momentum slices are studied; the flow coincides
with $\xi_{st}$ above. In the same work the timelike geodesic that
produces the imaginary part is identified in the Rindler picture as a
radial geodesic between the inner and outer Rindler horizons. It is also shown that this geodesic is not the set of
fixed points of the modular momentum flow, so that its role in the
replica picture was left open.

This paper takes these questions up for a timelike interval on the plane. We start from the symmetries of the theory, as in \cite{Jiang_2017}, and build three boundary modular flows as combinations of symmetry generators, together with their bulk extensions. One has the two spacelike branches as its fixed points, one has the timelike curve, and one belongs to a spacelike interval, whose associated spacelike extremal curve gives the corresponding holographic entanglement entropy. We keep this flow because it completes the picture in Rindler space. The zero sets of the first two flows are the spacelike and timelike extremal curves, and their lengths reproduce the TEE. We do this for CFT$_2$ and its AdS$_3$ dual, first in Poincar\'e and then in BMS coordinates, and afterwards for CCFT$_2$ and its flat-space dual. In the flat case the real and imaginary parts come out equal, $S^{(T)}=(1+i)\,l_u/(2Gl_\phi)$, and the flat limit of the AdS result recovers them once the length is taken between $r=0$ and the point $r_I$, following the prescription of \cite{Fareghbal:2024lqa}.

The second half is about geometry. In
\cite{shalamzari2026modularflowholographictimelike} we constructed the
boundary and bulk modular flows of spacelike and timelike intervals from
the symmetry generators of CFT$_2$ and CCFT$_2$, and found a second
Killing vector, $\xi_{tt}$, whose zero set is the timelike extremal
curve, together with the flat limit of these flows. Here we use these
flows in the Rindler picture. After the Rindler transformation, the flow
of the spacelike branches and the flow of a spacelike interval become
constant vector fields, and their extremal curves sit on the inner and
the outer horizon, respectively. Exchanging $l_u\leftrightarrow \ell
l_\phi$ swaps the two, which amounts to a move between horizons. The
timelike curve is another matter. Its Rindler frame location was
described in \cite{Wen2025}; relative to that work, we show that it is
the zero set of $\xi_{tt}$, a geodesic because it is the fixed-point set
of an isometry, and we fix it explicitly as $\tilde U=\tilde V=0$ with
$-T_{\tilde U}T_{\tilde V}<\tilde\rho<T_{\tilde U}T_{\tilde V}$, such that it leaves the inner horizon, touches the
outer horizon at its midpoint, and returns to the inner horizon. We show that no such curve
exists outside the region between the horizons, and we compute its
length directly in Rindler coordinates. Its length, $i\pi\ell$, depends
neither on the interval nor on the horizon radii, which explains why the
imaginary part is universal. In flat space the outer horizon is sent to
infinity. The spacelike flow now has its curve on the cosmological
horizon of an FSC, and the timelike curve runs from that horizon
outward, with the part contributing to the entropy ending at
$\tilde r_I$ and giving $2il_u/l_\phi$. As a consistency check,
carrying the point $I$ from AdS$_3$ through the Rindler coordinates to
the BMS gauge lands on the same $\tilde r_I$ as the direct flat-space
computation.

The paper is organized as follows. Section~\ref{sec21} constructs the modular flows of a timelike interval in CFT$_2$ on the plane and recovers the TEE from their extremal curves. The Rindler transformation and the identification of each curve with respect to the horizons in BTZ occupy section~\ref{sec22}. Section~\ref{sec31} follows the construction of section \ref{sec21} for CCFT$_2$ and relates the result to the flat limit. Section~\ref{sec32} pursues the geometrical interpretation of bulk extremal curves in the FSC, using the flat Rindler transformation. We close with a discussion in section~\ref{sec5}.

	\section{Holographic Timelike Entanglement Entropy in Rindler AdS}
	
	In this section we study timelike entanglement entropy (TEE) through the modular flow of a timelike interval, and then give its extremal curves a geometric meaning in Rindler space. We start with a CFT$_2$ on the plane. There we write the boundary and bulk modular flows in Poincar\'e and BMS coordinates, and use their fixed points to compute the entropy. The BMS form is the one we need, since the Rindler transformation acts on it directly. In the second subsection we move to Rindler BTZ. The extremal curve of one flow lies on the outer horizon, and that of another on the inner horizon. The timelike curve stretches from the inner horizon to the outer horizon.
	
	\subsection{TEE from Modular Flow for CFT$_{2}$ on the Plane}
	\label{sec21}
	
	Timelike entanglement entropy was introduced in \cite{Doi2022,Doi_2023} as the entropy of a timelike boundary subsystem. It follows from analytic continuation of the spacelike counterpart and, unlike the spacelike case, is complex. Holographically, the quantity is again tied to extremal curves in the bulk. For AdS$_3$/CFT$_2$, it was shown in \cite{Doi2022} that the real part is given by a two branch spacelike curve, which starts at the endpoints of the interval on the CFT$_2$ boundary and extends into the bulk. The two branches are joined inside the bulk by a timelike curve, and this curve gives rise to the imaginary part of the entropy.
	
	Extremal curves can also be found from modular flow. In \cite{shalamzari2026modularflowholographictimelike}, the symmetries of the theory were used to construct the modular flows of a timelike interval on the CFT$_2$ boundary. The fixed points of the bulk flow coincide with the extremal curves \cite{Apolo_2020}, which gave the curve equations needed to compute the timelike entanglement entropy.
	
	We apply the same method starting from a general timelike interval on the CFT$_2$ boundary, together with the corresponding boundary and bulk modular flows in Poincar\'e coordinates. A suitable coordinate transformation takes us to BMS coordinates, where we compute the timelike entanglement entropy. These coordinates will help later, when we look for the geometric picture in Rindler space.
	
	The AdS$_3$ metric in Poincar\'e coordinates reads
	\begin{equation}
		ds^{2} = \dfrac{\ell^{2}}{z^2} \left(- dt^2 + dx^{2} + dz^2\right),
		\label{1}
	\end{equation}
	where $\ell$ is the AdS radius and the dual CFT$_2$ lives on the conformal boundary at $z \to 0$.
	On the conformal boundary we take a timelike interval
	\begin{equation}
		-\dfrac{l_{x}}{2} < t < \dfrac{l_{x}}{2}, \qquad -\dfrac{l_{t}}{2} < x < \dfrac{l_{t}}{2},
	\end{equation}
	where the condition $l_{x} > l_{t}$ ensures that the interval is timelike. In \cite{shalamzari2026modularflowholographictimelike}, the boundary modular flows of the timelike interval were found to be
	
	\begin{equation}
		\begin{split}
			\xi_{st}^{b} ={}& \dfrac{\pi (l_{x}^{2} l_{t} - l_{t}^3 -8 l_{x} x t + 4 l_{t}(x^{2} + t^{2}))}{2 (l_{x}^2-l_{t}^2)} \partial_{t}  + \dfrac{\pi ( - l_{t}^{2} l_{x} + l_{x}^3 + 8 l_{t} x t - 4 l_{x}(x^{2} + t^{2} ))}{2 (l_{x}^2-l_{t}^2)} \partial_{x},
			\label{2}
		\end{split}
	\end{equation}
	\vspace{0.1cm}
	\begin{equation}
		\begin{split}
			\xi_{tt}^{b} = \dfrac{\pi (-l_{x}^{2} l_{t}+l_{t}^3 - 8 l_{x} x t + 4 l_{t}(x^2 + t^2 ))}{2 (l_{x}^2-l_{t}^2)} \partial_{t}  + \dfrac{\pi (l_{t}^{2} l_{x}-l_{x}^3 + 8 l_{t} x t - 4 l_{x}(x^2 + t^2))}{2 (l_{x}^2-l_{t}^2)} \partial_{x}.
			\label{4} 
		\end{split}
	\end{equation}
	We also consider the modular flow $\xi_{ss}$. It corresponds to a spacelike interval on the boundary~\cite{Fareghbal:2020dtq} and therefore does not enter the computation of the timelike entanglement entropy. In Rindler space, however, it will help us build a fuller picture of the timelike curve. Its boundary form is
	\begin{equation}
		\begin{split}
			\xi_{ss}^{b} = \dfrac{\pi (-l_{t}^{2} l_{x}+l_{x}^3+8 l_{t} x t - 4 l_{x}(x^{2} + t^{2} ))}{2 (l_{x}^2-l_{t}^2)} \partial_{t} + \dfrac{\pi (l_{x}^{2} l_{t}-l_{t}^3-8 l_{x} x t + 4 l_{t}(x^{2} + t^{2} ))}{2 (l_{x}^2-l_{t}^2)} \partial_{x}.
		\end{split}
	\end{equation}
	The bulk extensions of these boundary flows are given by
	\begin{equation}
		\begin{split}
			\xi_{st}^{B} ={}& \dfrac{\pi (-l_{x}^{2} l_{t}+l_{t}^3+8 l_{x} x t - 4 l_{t}(x^{2} + t^{2} + z^{2}))}{2 (l_{t}^2-l_{x}^2)} \partial_{t} \\
			&+ \dfrac{\pi (l_{t}^{2} l_{x}-l_{x}^3-8 l_{t} x t + 4 l_{x}(x^{2} + t^{2} - z^{2}))}{2 (l_{t}^2-l_{x}^2)} \partial_{x} \\
			&- \dfrac{4 \pi (- l_{x} x+l_{t} t) z}{l_{t}^{2}-l_{x}^2} \partial_{z},
		\end{split}
		\label{37}
	\end{equation}
	
	\begin{equation}
		\begin{split}
			\xi_{tt}^{B} ={}& \dfrac{\pi (-l_{x}^{2} l_{t}+l_{t}^3 - 8 l_{x} x t + 4 l_{t}(x^{2} + t^{2} + z^{2}))}{2 (l_{x}^2-l_{t}^2)} \partial_{t} \\
			&+ \dfrac{\pi (l_{t}^{2} l_{x}-l_{x}^3 + 8 l_{t} x t - 4 l_{x}(x^{2} + t^{2} - z^{2}))}{2 (l_{x}^2-l_{t}^2)} \partial_{x} \\
			&- \dfrac{4 \pi ( l_{x} x - l_{t} t) z}{l_{x}^{2}-l_{t}^2} \partial_{z},
		\end{split}
		\label{48}
	\end{equation}

	\begin{equation}
		\begin{split}
			\xi_{ss}^{B} ={}& \dfrac{\pi (-l_{t}^{2} l_{x}+l_{x}^3+8 l_{t} x t - 4 l_{x}(x^{2} + t^{2} + z^{2}))}{2 (l_{x}^2-l_{t}^2)} \partial_{t} \\
			&+ \dfrac{\pi (l_{x}^{2} l_{t}-l_{t}^3-8 l_{x} x t + 4 l_{t}(x^{2} + t^{2} - z^{2}))}{2 (l_{x}^2-l_{t}^2)} \partial_{x} \\
			&- \dfrac{4 \pi (- l_{t} x+l_{x} t) z}{l_{x}^{2}-l_{t}^2} \partial_{z}.
		\end{split}
		\label{38}
	\end{equation}
	
	By comparing the bulk flows $\xi_{st}^{B}$ in \eqref{37} and $\xi_{ss}^{B}$ in \eqref{38}, we find that these are exchanged under
	\begin{equation}
		l_x \longleftrightarrow l_t.
		\label{39}
	\end{equation}
	The Poincar\'e coordinates are related to the BMS ones by
	\begin{equation}
		t = u - \dfrac{\ell^{2}}{r}, \qquad x = \ell \phi, \qquad z = \dfrac{\ell^{2}}{r},
		\label{47}
	\end{equation}
	which brings the metric in \eqref{1} into the form of 
	\begin{equation}
		ds^{2} = - \dfrac{r^{2}}{\ell^2} \: du^{2} - 2 du dr + r^{2} d\phi^{2},
		\label{10}
	\end{equation}
	with $\phi$ non-compact.
	
	In these coordinates, we consider the boundary interval
	\begin{equation}
		-\dfrac{\ell l_{\phi}}{2} < u < \dfrac{\ell l_{\phi}}{2}, \qquad -\dfrac{l_{u}}{2 \ell} < \phi < \dfrac{l_{u}}{2 \ell}.
	\end{equation}
	The constants $l_{\phi}$ and $l_{u}$ are related to $l_{x}$ and $l_{t}$ through
	\begin{equation}
		\ell l_{\phi} = l_{x}, \qquad l_{u} = l_{t},
		\label{28}
	\end{equation}
	so the condition $l_{x} > l_{t}$ implies $\ell l_{\phi} > l_{u}$, which ensures that the interval is timelike.
	
	Under the transformations \eqref{47}, the bulk modular flows \eqref{37}--\eqref{38} take the form
	
\begin{figure}
		\centering
		\includegraphics[width=0.45\linewidth]{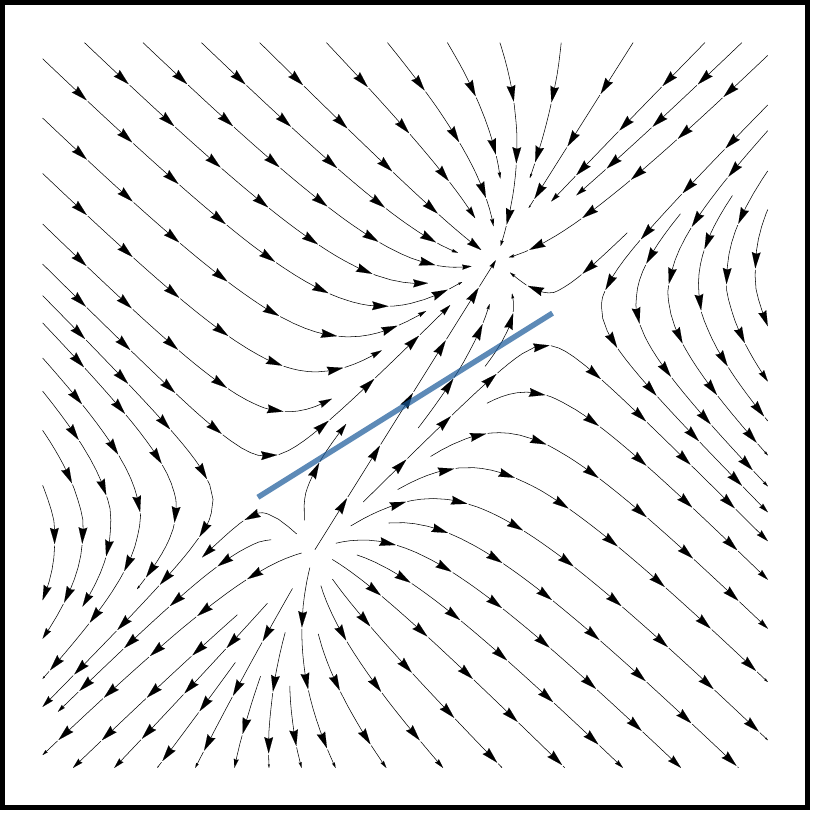}\\
		(a)\\[1.5ex]
		\begin{minipage}{0.48\linewidth}
			\centering
			\includegraphics[width=\linewidth]{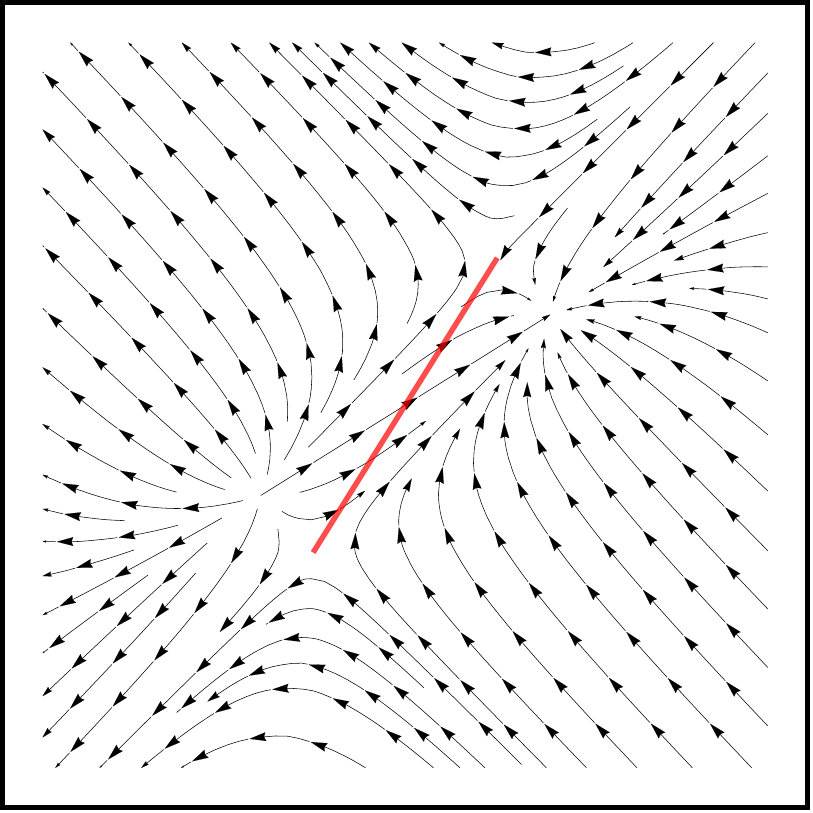}\\
			(b)
		\end{minipage}
		\hfill
		\begin{minipage}{0.48\linewidth}
			\centering
			\includegraphics[width=\linewidth]{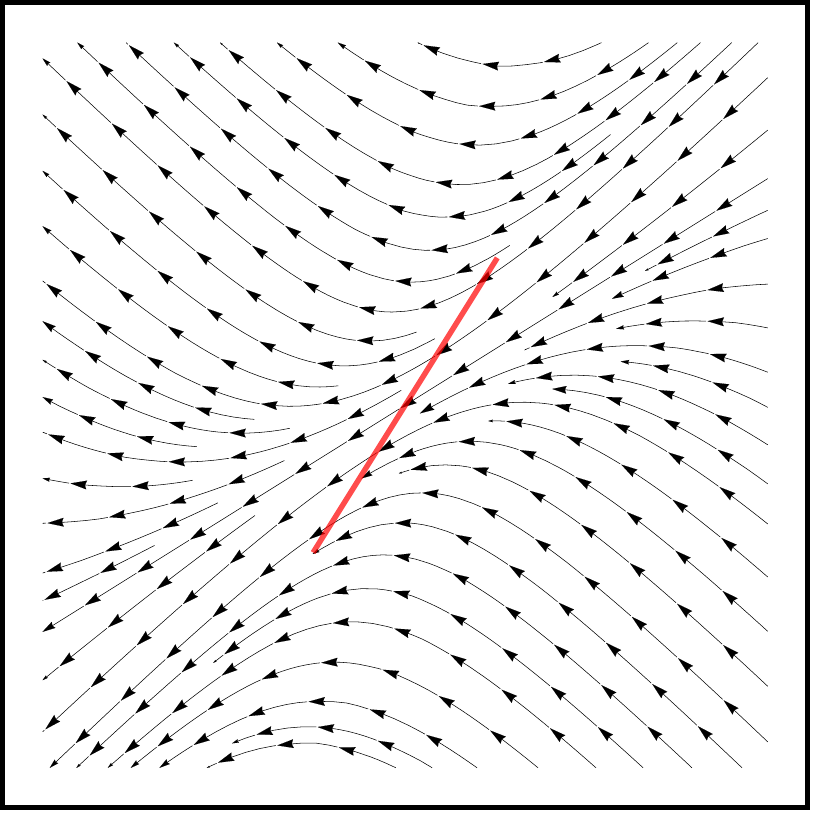}\\
			(c)
		\end{minipage}
		\caption{ \justifying Boundary modular flows of AdS$_3$ on the plane for (a) $\xi^{b}_{ss}$, (b) $\xi^{b}_{st}$ and (c) $\xi^{b}_{tt}$. The spacelike interval is shown in blue and the timelike interval in red.}
		\label{102}
	\end{figure}

	\begin{widetext}
		\begin{equation}
			\begin{split}
				\xi^{B}_{st} ={}& \dfrac{\pi \left(l_{u} \ell^2 l_{\phi}^{2} - l_{u}^{3} - 8 l_{\phi} \ell^{2} u \phi + 4 l_{u} \left(u^{2} +\ell^{2} \phi^{2}\right)\right)}{2 \left(\ell^2 l_{\phi}^{2} - l_u^{2}\right)} \partial_{u} + \dfrac{4 \pi \left(l_{u} \left(\ell^{2} - r u\right) + \ell^{2} l_{\phi} r \phi\right) }{\ell^{2} l_{\phi}^{2} - l_{u}^{2}} \partial_{r} \\
				&+ \dfrac{\pi \left(l_{\phi}^{3} \ell^{2} r - l_{u}^{2} l_{\phi} r - 4 l_{\phi} r u^{2} - 8 l_{u} \phi \left(\ell^{2} - r u\right) + \ell^{2} l_{\phi} \left(8 u - 4 r \phi^{2} \right)\right)}{2 r \left(\ell^{2} l_{\phi}^{2} - l_{u}^{2}\right)} \partial_{\phi},
			\end{split}
			\label{6}
		\end{equation}
		\begin{equation}
			\begin{split}
				\xi^{B}_{tt} ={}& \dfrac{\pi \left(-l_{u} \ell^2 l_{\phi}^{2} + l_{u}^{3} - 8 l_{\phi} \ell^{2} u \phi + 4 l_{u} \left(u^{2} +\ell^{2} \phi^{2}\right)\right)}{2 \left(\ell^2 l_{\phi}^{2} - l_u^{2}\right)} \partial_{u} + \dfrac{4 \pi \left(l_{u} \left(\ell^{2} - r u\right) + \ell^{2} l_{\phi} r \phi\right) }{\ell^{2} l_{\phi}^{2} - l_{u}^{2}} \partial_{r} \\
				&+ \dfrac{\pi \left(-l_{\phi}^{3} \ell^{2} r + l_{u}^{2} l_{\phi} r - 4 l_{\phi} r u^{2} - 8 l_{u} \phi \left(\ell^{2} - r u\right) + \ell^{2} l_{\phi} \left(8 u - 4 r \phi^{2} \right)\right)}{2 r \left(\ell^{2} l_{\phi}^{2} - l_{u}^{2}\right)} \partial_{\phi},
			\end{split}
			\label{7}
		\end{equation}
		\begin{equation}
			\begin{split}
				\xi^{B}_{ss} ={}& - \dfrac{\pi \ell \left(l_{u}^2  l_{\phi} - \ell^2 l_{\phi}^{3} - 8 l_{u}  u \phi + 4 l_{\phi} \left(u^{2} +\ell^{2} \phi^{2}\right)\right)}{2 \left(\ell^2 l_{\phi}^{2} - l_u^{2}\right)} \partial_{u} - \dfrac{4 \pi \ell \left(l_\phi \left(\ell^2-r u \right) +l_u r \phi\right)}{\ell^2 l_\phi^2 - l_u^2} \partial_r \\
				&+ \dfrac{\pi \left(-l_{u}^{3} r  + 8 \ell^2 l_{\phi} \left(\ell^2 - ru\right) \phi  +l_u \left(\ell^2 l_\phi^2 r+ 4 r u^2 + \ell^2  \left(4r \phi^2 - 8 u\right)\right) \right)}{2 \ell \left(\ell^{2} l_{\phi}^{2} - l_{u}^{2}\right) r} \partial_{\phi}.
			\end{split}
			\label{23}
		\end{equation}
	\end{widetext}
	As we saw earlier in \eqref{39}, the flows $\xi_{st}^B$ and $\xi_{ss}^B$ are related in these coordinates as well, under the replacement \cite{Fareghbal:2020dtq}
	\begin{equation}
		l_u \longrightarrow \ell l_\phi, \qquad l_\phi \longrightarrow \dfrac{l_u}{\ell}.
		\label{40}
	\end{equation}
	The corresponding boundary modular flows are obtained by taking $r\rightarrow \infty$, and reduce to (see Fig.~\ref{102})
	
	\begin{equation}
		\begin{split}
			\xi^{b}_{st} =  \dfrac{\pi \left(l_{u} \ell^2 l_{\phi}^{2} - l_{u}^{3} - 8 l_{\phi} \ell^{2} u \phi + 4 l_{u} \left(u^{2} +\ell^{2} \phi^{2}\right)\right)}{2 \left(\ell^2 l_{\phi}^{2} - l_u^{2}\right)} \partial_{u} \\+ \dfrac{\pi \left(l_{\phi}^{3} \ell^{2}  - l_{u}^{2} l_{\phi}  - 4 l_{\phi}  u^{2} + 8 l_{u} \phi u - 4 \ell^{2} l_{\phi} \phi^{2} \right)}{2 \left(\ell^{2} l_{\phi}^{2} - l_{u}^{2}\right)} \partial_{\phi} 
			\label{11}
		\end{split}
	\end{equation}
	\begin{equation}
		\begin{split}
			\xi^{b}_{tt} =  \dfrac{\pi \left(-l_{u} \ell^2 l_{\phi}^{2} + l_{u}^{3} - 8 l_{\phi} \ell^{2} u \phi + 4 l_{u} \left(u^{2} +\ell^{2} \phi^{2}\right)\right)}{2 \left(\ell^2 l_{\phi}^{2} - l_u^{2}\right)} \partial_{u} \\ + \dfrac{\pi \left(- l_{\phi}^{3} \ell^{2}  + l_{u}^{2} l_{\phi}  - 4 l_{\phi}  u^{2} + 8 l_{u} \phi u - 4 \ell^{2} l_{\phi} \phi^{2} \right)}{2 \left(\ell^{2} l_{\phi}^{2} - l_{u}^{2}\right)} \partial_{\phi}
			\label{49}
		\end{split}
	\end{equation}
	\begin{equation}
		\begin{split}
			\xi^{b}_{ss} = - \dfrac{\pi \ell \left(l_{u}^2  l_{\phi} - \ell^2 l_{\phi}^{3} - 8 l_{u}  u \phi + 4 l_{\phi} \left(u^{2} +\ell^{2} \phi^{2}\right)\right)}{2 \left(\ell^2 l_{\phi}^{2} - l_u^{2}\right)} \partial_{u} \\ - 	\dfrac{\pi \left(l_{u}^{3}  - \ell^2 l_{u} l_{\phi}^2  - 4 l_{u}  u^{2} + 8 \ell^2 l_{\phi} u \phi  - 4 \ell^{2} l_{u} \phi^{2} \right)}{2 \ell \left(\ell^{2} l_{\phi}^{2} - l_{u}^{2}\right)} \partial_{\phi} 
			\label{}
		\end{split}
	\end{equation}
	
	Setting the components of the modular flow $\xi^{B}_{st}$ to zero gives
	the equations of the spacelike extremal curve,
	\begin{equation*}
		l_u \left(\ell^2 -ru\right) + \ell^2 l_\phi r \phi=0,
	\end{equation*}
	\begin{equation}
		l_{u} \ell^2 l_{\phi}^{2} - l_{u}^{3} - 8 l_{\phi} \ell^{2} u \phi + 4 l_{u} \left(u^{2} +\ell^{2} \phi^{2}\right)=0,
		\label{24}
	\end{equation}
	similarly, the modular flow $\xi^{B}_{tt}$ in \eqref{7} yields the
	timelike extremal curve,
	\begin{equation*}
		l_{u} \left(\ell^{2} - r u\right) + \ell^{2} l_{\phi} r \phi=0,
	\end{equation*}
	\begin{equation}
		-l_{u} \ell^2 l_{\phi}^{2} + l_{u}^{3} - 8 l_{\phi} \ell^{2} u \phi + 4 l_{u} \left(u^{2} +\ell^{2} \phi^{2}\right)=0.
		\label{25}
	\end{equation}
	For the extremal curve associated with $\xi^{B}_{ss}$, see~\cite{Fareghbal:2020dtq}.

	The bulk modular flow $\xi^{B}_{st}$ vanishes on a two-branch spacelike curve \cite{shalamzari2026modularflowholographictimelike,Wen2025}. These branches are anchored at the endpoints of the boundary interval and extend into the bulk. A timelike curve joins the two branches, and $\xi^{B}_{tt}$ vanishes on it \cite{shalamzari2026modularflowholographictimelike}.
	
	Substituting the spacelike extremal curve \eqref{24} into the metric \eqref{10}, the line element becomes
	\begin{equation}
		ds_{\xi_{st}} = \dfrac{\ell \: dr}{\sqrt{r^{2} + \dfrac{4 \ell^{4}}{\ell^{2} l_{\phi}^2 - l_u^{2}}}}.
		\label{18}
	\end{equation}
	Each branch of the spacelike extremal curve runs from $r=0$ out to the asymptotic boundary, $r \to \infty$, and the length integral diverges there. To regulate this divergence, we introduce a cutoff at $r_{\epsilon}=\ell^{2}/\epsilon$, with $\epsilon \to 0$. The length of the spacelike extremal curve is therefore twice the contribution of a single branch:
	\begin{equation}
		\begin{split}
			L_{st} &= 2 \int_{0}^{r_\epsilon} ds_{\xi_{st}} = 2 \ell \sinh^{-1}\left(\dfrac{\sqrt{\ell^{2} l_{\phi}^{2} - l_{u}^{2}}}{2 \epsilon}\right) = 2 \ell \log \left(\dfrac{ \sqrt{\ell^{2}l_{\phi}^{2} - l_{u}^{2}}}{\epsilon}\right).
		\end{split}
	\end{equation}
	Now, we turn to the timelike extremal curve. By combining the equations in \eqref{25}, we get
	\begin{equation}
		\phi^{2} = -\dfrac{l_{u}^{2}}{4 \ell^{2}} + \dfrac{\ell^{2} l_{u}^{2}}{r^{2}\left(\ell^{2} l_{\phi}^{2} - l_{u}^{2}\right)}.
		\label{26}
	\end{equation}
	This relation shows that the radial coordinate attains a maximum, given by
	\begin{equation}
		r_{\max} = \dfrac{2 \ell^{2}}{\sqrt{\ell^{2} l_{\phi}^{2} - l_{u}^{2}}}.
	\end{equation}
	Hence the timelike curve stays within the bulk and never reaches the asymptotic boundary.
	
	The line element along the timelike curve is
	\begin{equation}
		ds_{\xi_{tt}} = -\dfrac{\ell \: dr}{\sqrt{r^{2} - \dfrac{4 \ell^{4}}{\ell^{2} l_{\phi}^2 - l_u^{2}}}},
		\label{19}
	\end{equation}
	hence,
	\begin{equation}
		L_{tt} = 2\int_{0}^{r_{\max}} ds_{\xi_{tt}} = i \pi \ell.
	\end{equation}
	The total length associated with the timelike interval is obtained by summing the contributions from the spacelike and timelike extremal curves. The holographic timelike entanglement entropy for CFT$_2$ on the plane is derived as follows
	
	\begin{equation}
		S^{(T)} = \dfrac{L_{st} + L_{tt}}{4G_{N}}= \dfrac{\ell}{2 G_{N}}  \log \left(\dfrac{ \sqrt{\ell^{2}l_{\phi}^{2} - l_{u}^{2}}}{\epsilon}\right) + \dfrac{i \pi \ell}{4 G_{N}},
		\label{17}
	\end{equation}
	using the central charge relation $c= 3 \ell/\left(2 G_{N}\right)$, it can be rewritten as
	\begin{equation}
		S^{(T)} = \dfrac{c}{3}  \log \left(\dfrac{ \sqrt{\ell^{2}l_{\phi}^{2} - l_{u}^{2}}}{\epsilon}\right) + \dfrac{i  \pi c}{6}.
	\end{equation}
	Setting $\ell l_{\phi} = \Delta t$ and $l_{u} = \Delta x$ reproduces the result of \cite{Doi_2023}.

	\subsection{Holographic Timelike Entanglement Entropy in Rindler BTZ}
	\label{sec22}
	In Section~\ref{sec21} we obtained the boundary and bulk modular flows $\xi^B_{st}$, $\xi^B_{ss}$ and $\xi^B_{tt}$ for a timelike interval on the plane, in BMS coordinates. Their extremal curves admit a geometric interpretation via the Rindler method. Under the Rindler transformation of \cite{Fareghbal:2024lqa}, the extremal curve of $\xi^B_{ss}$ lies on the outer horizon of a BTZ black hole. In the flat-space limit this horizon is sent to infinity, so neither the flow nor its extremal curve has a finite limit. The same transformation places the extremal curve of $\xi^B_{st}$ on the inner horizon \cite{Fareghbal:2020dtq}, which after taking the flat limit maps to the cosmological horizon of the flat-space cosmology (FSC); consequently $\xi^B_{st}$ and its extremal curve survive the flat-space limit.
	
	We also saw that $\xi^B_{ss}$ and $\xi^B_{st}$ are exchanged under $l_x\leftrightarrow l_t$ in Poincaré coordinates, equivalently under $l_u\to \ell\, l_\phi$, $l_\phi\to l_u/\ell$ in BMS coordinates. This exchange corresponds to moving between the inner and outer horizons of the BTZ black hole.
	
	It remains to interpret $\xi^B_{tt}$. The imaginary part of the timelike entanglement entropy arises from a timelike curve, namely the set of fixed points of $\xi^B_{tt}$. To give this flow and its timelike curve a geometric description, we again turn to Rindler space.
	
	Consider the coordinate transformation 
	\begin{equation}
		u=\ell^2\left( \sqrt{\dfrac{2}{\rho}}+\left(U-V\right) \right), \quad r=\sqrt{\dfrac{\rho}{2}}, \quad \phi=\ell\left(U+V\right),
		\label{29}
	\end{equation}
	which brings the metric \eqref{10} to
	\begin{equation}
		ds^2=\ell^2\left(\dfrac{d\rho^2}{4\rho^2} + 2 \rho dU dV\right).
		\label{31}
	\end{equation}
	The interval parameters $l_u$ and $\ell l_\phi$ of \eqref{28} are then expressed through
	$l_U$ and $l_V$, the coordinate separations of the endpoints in $U$ and $V$, as
	
	\begin{equation}
		l_u=\ell^2 \left(l_U-l_V\right), \qquad \ell l_\phi= \ell^2 \left(l_U+l_V\right).
		\label{54}
	\end{equation}
	Here $\ell$ and $l_U$ are positive constants, and the endpoints of the interval sit at $(U,V)=(\pm l_U/2,\,\mp l_V/2)$. The interval is timelike, $\ell l_\phi>l_u$, when $l_V>0$.

	In the coordinates $(\rho,U,V)$ the modular flows of \eqref{6}--\eqref{23} read
	\begin{equation}
		\begin{split}
			\xi_{st}^{B} = 4 \pi \rho \left(\dfrac{U}{l_U} + \dfrac{V}{l_V}\right) \partial_{\rho} +\left(\dfrac{\pi}{l_V \rho} -\dfrac{2\pi U^2}{l_U} + \dfrac{\pi l_U}{2}\right) \partial_{U} +\left(\dfrac{\pi}{l_U \rho} -\dfrac{2\pi V^2}{l_V} + \dfrac{\pi l_V}{2}\right) \partial_{V},
		\end{split}
		\label{41}
	\end{equation}
	\vspace{0.1cm}
	\begin{equation}
		\begin{split}
			\xi_{tt}^{B} = 4 \pi \rho \left(\dfrac{U}{l_U} + \dfrac{V}{l_V}\right) \partial_{\rho} +\left(\dfrac{\pi}{l_V \rho} -\dfrac{2\pi U^2}{l_U} - \dfrac{\pi l_U}{2}\right) \partial_{U} +\left(\dfrac{\pi}{l_U \rho} -\dfrac{2\pi V^2}{l_V} - \dfrac{\pi l_V}{2}\right) \partial_{V},
		\end{split}
		\label{43}
	\end{equation}
		\vspace{0.1cm}
	\begin{equation}
		\begin{split}
			\xi_{ss}^{B} = 4 \pi \rho \left(\dfrac{U}{l_U} - \dfrac{V}{l_V}\right) \partial_{\rho} +\left(-\dfrac{\pi}{l_V \rho} -\dfrac{2\pi U^2}{l_U} + \dfrac{\pi l_U}{2}\right) \partial_{U} +\left(\dfrac{\pi}{l_U \rho} +\dfrac{2\pi V^2}{l_V} - \dfrac{\pi l_V}{2}\right) \partial_{V}.
		\end{split}
		\label{42}
	\end{equation}
	The extremal curve of each flow is its set of fixed points. Only two of the three
	component equations are independent, and solving them gives
	\begin{subequations}\label{30}
		\begin{align}
			\xi^B_{st}:&\quad l_VU+l_UV=0,\quad \rho\,(4U^2-l_U^2)=\frac{2l_U}{l_V},\label{30a}\\[1.5ex]
			\xi^B_{tt}:&\quad l_VU+l_UV=0,\quad \rho\,(l_U^2+4U^2)=\frac{2l_U}{l_V},\label{30b}\\[1.5ex]
			\xi^B_{ss}:&\quad l_VU-l_UV=0,\quad \rho\,(l_U^2-4U^2)=\frac{2l_U}{l_V}.\label{30c}
		\end{align}
	\end{subequations}
	The two flows 	$\xi^B_{ss}$ and $\xi^B_{st}$ are exchanged by $l_V\to-l_V$ \cite{Wen2025}, which is the swap
	$l_u\leftrightarrow \ell l_\phi$ of the previous section. Since $l_U/l_V>0$, the right-hand sides are positive. The curve for $\xi^B_{tt}$ therefore exists for all $U$, while those for $\xi^B_{st}$ and $\xi^B_{ss}$ exist only for $4U^2>l_U^2$ and $4U^2<l_U^2$, respectively.
	
	To identify each flow and its extremal curve geometrically, we pass to Rindler coordinates. The Rindler transformation is valid in the region between the two horizons of the BTZ black hole. This is also the only region where the timelike extremal curve exists, as shown in Appendix~\ref{appendix}. The transformation is given by
	\begin{widetext}
		\begin{equation}
			\begin{split}
				\rho &= \left(\sqrt{\dfrac{T_{\tilde{U}} T_{\tilde{V}} + \tilde{\rho}}{l_U l_V T_{\tilde{U}} T_{\tilde{V}} }} \cosh\left(T_{\tilde{U}} \tilde{U} + T_{\tilde{V}} \tilde{V}\right) - \sqrt{\dfrac{T_{\tilde{U}} T_{\tilde{V}} - \tilde{\rho}}{l_U l_V T_{\tilde{U}} T_{\tilde{V}} }} \sinh\left(T_{\tilde{U}} \tilde{U} - T_{\tilde{V}} \tilde{V}\right) \right)^2, \\[3ex]
				U &= \dfrac{\sqrt{\dfrac{l_U \left(T_{\tilde{U}} T_{\tilde{V}} + \tilde{\rho}\right)}{4 l_V T_{\tilde{U}} T_{\tilde{V}}}} \sinh\left(T_{\tilde{U}} \tilde{U} + T_{\tilde{V}} \tilde{V}\right)  - \sqrt{\dfrac{l_U \left(T_{\tilde{U}} T_{\tilde{V}} - \tilde{\rho}\right)}{4 l_V T_{\tilde{U}} T_{\tilde{V}}}} \cosh\left(T_{\tilde{U}} \tilde{U} - T_{\tilde{V}} \tilde{V}\right)}{\sqrt{\dfrac{T_{\tilde{U}} T_{\tilde{V}} + \tilde{\rho}}{l_U l_V T_{\tilde{U}} T_{\tilde{V}} }} \cosh\left(T_{\tilde{U}} \tilde{U} + T_{\tilde{V}} \tilde{V}\right) - \sqrt{\dfrac{T_{\tilde{U}} T_{\tilde{V}} - \tilde{\rho}}{l_U l_V T_{\tilde{U}} T_{\tilde{V}} }} \sinh\left(T_{\tilde{U}} \tilde{U} - T_{\tilde{V}} \tilde{V}\right) }, \\[3ex]
				V &= \dfrac{\sqrt{\dfrac{l_V \left(T_{\tilde{U}} T_{\tilde{V}} + \tilde{\rho}\right)}{4 l_U T_{\tilde{U}} T_{\tilde{V}}}} \sinh\left(T_{\tilde{U}} \tilde{U} + T_{\tilde{V}} \tilde{V}\right)  + \sqrt{\dfrac{l_V \left(T_{\tilde{U}} T_{\tilde{V}} - \tilde{\rho}\right)}{4 l_U T_{\tilde{U}} T_{\tilde{V}}}} \cosh \left(T_{\tilde{U}} \tilde{U} - T_{\tilde{V}} \tilde{V}\right)}{\sqrt{\dfrac{T_{\tilde{U}} T_{\tilde{V}} + \tilde{\rho}}{l_U l_V T_{\tilde{U}} T_{\tilde{V}} }} \cosh\left(T_{\tilde{U}} \tilde{U} + T_{\tilde{V}} \tilde{V}\right) - \sqrt{\dfrac{T_{\tilde{U}} T_{\tilde{V}} - \tilde{\rho}}{l_U l_V T_{\tilde{U}} T_{\tilde{V}} }} \sinh\left(T_{\tilde{U}} \tilde{U} - T_{\tilde{V}} \tilde{V}\right) }.
			\end{split}
			\label{35}
		\end{equation}
	\end{widetext}
	
	The inverse transformations are
	\begin{widetext}
		\begin{equation}
			\begin{split}
				\tilde{U} &= \dfrac{\log \left[\dfrac{\left(-2+ \left(l_U + 2 U\right) \rho \left(l_V - 2V\right)\right) \left(2+ \left(l_U+2U\right) \rho \left(l_V+2V\right)\right)}{\left(2- \left(l_U - 2 U\right) \rho \left(l_V + 2V\right)\right) \left(2+ \left(l_U-2U\right) \rho \left(l_V-2V\right)\right)}\right]}{4 T_{\tilde{U}}}, \\[3ex]
				\tilde{V} &= \dfrac{\log \left[\dfrac{\left(2- \left(l_U - 2 U\right) \rho \left(l_V + 2V\right)\right) \left(2+ \left(l_U+2U\right) \rho \left(l_V+2V\right)\right)}{\left(2+ \left(l_U - 2 U\right) \rho \left(l_V - 2V\right)\right) \left(-2+ \left(l_U+2U\right) \rho \left(l_V-2V\right)\right)}\right]}{4 T_{\tilde{V}}}, \\[3ex]
				\tilde{\rho} &= \dfrac{T_{\tilde{U}} T_{\tilde{V}} \left(4 +16 U V \rho  + \left(l_U^2 - 4 U^2\right) \left(l_V^2 - 4 V^2\right) \rho^2\right)}{4 l_U l_V \rho},
			\end{split}
			\label{46}
		\end{equation}
	\end{widetext}
	where $T_{\tilde{U}}$ and $T_{\tilde{V}}$ are constants related to the outer horizon $\tilde{r}_{+}$ and inner horizon $\tilde{r}_{-}$, and are given by
	\begin{equation}
		T_{\tilde{U}} = \tilde{r}_{+} + \tilde{r}_{-}, \qquad  	T_{\tilde{V}} = \tilde{r}_{+} - \tilde{r}_{-}.
		\label{56}
	\end{equation}
	
	In these coordinates the metric \eqref{31} takes the form
	\begin{equation}
		\begin{split}
			ds^2 = \ell^2\left(T_{\tilde{U}}^2 d\tilde{U}^2 + 2\tilde{\rho}  d\tilde{U} d\tilde{V} + T_{\tilde{V}}^2  d\tilde{V}^2
			+ \dfrac{ d\tilde{\rho}^2}{4 \left(\tilde{\rho}^2 - T_{\tilde{U}}^2T_{\tilde{V}}^2\right)}\right) .
		\end{split}
		\label{32}
	\end{equation}
	This metric has the form of the rotating BTZ black hole. The horizons are found where the metric degenerates, that is, at $\tilde{\rho}^2=T_{\tilde{U}}^2T_{\tilde{V}}^2$. The outer horizon sits at $\tilde{\rho}=T_{\tilde{U}}T_{\tilde{V}}$ and the inner one at $\tilde{\rho}=-T_{\tilde{U}}T_{\tilde{V}}$. The region between them is  $-T_{\tilde{U}}T_{\tilde{V}}<\tilde{\rho}<T_{\tilde{U}}T_{\tilde{V}}$.
	
	The modular flows \eqref{41}--\eqref{42}, after the Rindler transformations, are
	\begin{equation}
		\hspace{-8.25cm}
		\xi_{st}^{B}= \dfrac{\pi}{T_{\tilde{U}}} \partial_{\tilde{U}} + \dfrac{\pi}{T_{\tilde{V}}} \partial_{\tilde{V}}.
	\end{equation}
	\vspace{-1.5cm}
	\begin{widetext}
		\begin{equation}
			\begin{split}
				\xi_{tt}^{B} ={}& -4\pi \sqrt{T_{\tilde{U}}^2 T_{\tilde{V}}^2-\tilde{\rho}^2}\, \sinh\!\left(T_{\tilde{U}} \tilde{U}-T_{\tilde{V}} \tilde{V}\right) \sinh\!\left(T_{\tilde{U}} \tilde{U}+T_{\tilde{V}} \tilde{V}\right) \partial_{\tilde{\rho}} \\
				& -\dfrac{\pi\left(\tilde{\rho}\, \sinh\!\left(2 T_{\tilde{U}} \tilde{U}\right)+T_{\tilde{U}} T_{\tilde{V}} \sinh\!\left(2 T_{\tilde{V}} \tilde{V}\right)\right)}{T_{\tilde{U}} \sqrt{T_{\tilde{U}}^2 T_{\tilde{V}}^2-\tilde{\rho}^2}}\, \partial_{\tilde{U}} \\
				& +\dfrac{\pi\left(T_{\tilde{U}} T_{\tilde{V}} \sinh\!\left(2 T_{\tilde{U}} \tilde{U}\right)+\tilde{\rho}\, \sinh\!\left(2 T_{\tilde{V}} \tilde{V}\right)\right)}{T_{\tilde{V}} \sqrt{T_{\tilde{U}}^2 T_{\tilde{V}}^2-\tilde{\rho}^2}}\, \partial_{\tilde{V}}.
			\end{split}
		\end{equation}
	\end{widetext}
	\begin{equation}
		\hspace{-8.25cm}
		\xi_{ss}^{B}= \dfrac{\pi}{T_{\tilde{U}}} \partial_{\tilde{U}} - \dfrac{\pi}{T_{\tilde{V}}} \partial_{\tilde{V}}.
	\end{equation}
	
	Both $\xi^B_{st}$ and $\xi^B_{ss}$ become constant vector fields, so their components never vanish in these coordinates. Since the norm of a vector vanishes wherever the vector does, the fixed points must lie where the norm is zero. Using the metric \eqref{32}, we obtain
	\begin{equation}
		\xi_{st}^2=2\pi^2\ell^2\left(1+\frac{\tilde\rho}{T_{\tilde U}T_{\tilde V}}\right),\qquad
		\xi_{ss}^2=2\pi^2\ell^2\left(1-\frac{\tilde\rho}{T_{\tilde U}T_{\tilde V}}\right).
	\end{equation}
	These vanish at $\tilde\rho=-T_{\tilde U}T_{\tilde V}$ and $\tilde\rho=+T_{\tilde U}T_{\tilde V}$, respectively, i.e. on the inner and outer horizons. The Rindler coordinates are singular there, which is why the flows appear nonvanishing; the fixed-point equations in $(\rho,U,V)$ confirm that the extremal curves of $\xi^B_{st}$ and $\xi^B_{ss}$ sit on these horizons.

	To find the timelike extremal curve, we set the components of  $\xi_{tt}^{B}$ to zero, which leads to 
	\begin{equation*}
		\sqrt{T_{\tilde{U}}^2 T_{\tilde{V}}^2-\tilde{\rho}^2}\, \sinh\!\left(T_{\tilde{U}} \tilde{U}-T_{\tilde{V}} \tilde{V}\right) \sinh\!\left(T_{\tilde{U}} \tilde{U}+T_{\tilde{V}} \tilde{V}\right)=0,
	\end{equation*}
	\begin{equation}
		\tilde{\rho}\, \sinh\!\left(2 T_{\tilde{U}} \tilde{U}\right)+T_{\tilde{U}} T_{\tilde{V}} \sinh\!\left(2 T_{\tilde{V}} \tilde{V}\right) =0.
	\end{equation}
	Solving these equations gives the equations of the timelike extremal curve,
	\begin{equation}
		\tilde U=0, \qquad \tilde V=0, \qquad -T_{\tilde U}T_{\tilde V}<\tilde\rho<T_{\tilde U}T_{\tilde V},
		\label{45}
	\end{equation}
	which is a curve between the two horizons.
	
	The timelike curve \eqref{45} is the line $\tilde U=\tilde V=0$, which runs along the
	$\tilde\rho$ direction. It is a geodesic of \eqref{32} for the following reason. The
	flow $\xi^B_{tt}$ is a Killing vector field of \eqref{32}, and its extremal curve is its
	set of zeros. This set is the fixed-point set of the one-parameter group of isometries
	generated by $\xi^B_{tt}$, and the fixed-point set of an isometry is totally geodesic.
	Hence the curve \eqref{45} is a geodesic. It is timelike because, on it, the metric
	\eqref{32} reduces to
	\begin{equation}
		ds^2=-\frac{\ell^2\,d\tilde\rho^2}{4\left(T_{\tilde U}^2T_{\tilde V}^2-\tilde\rho^2\right)},
	\end{equation}
	which is negative between the horizons, where the denominator is positive.
	
	The components of $\xi_{tt}^{B}$ at the horizons generally diverge. Using the metric \eqref{32}, we obtain the norm of the flow
	\begin{equation}
		\begin{split}
			\left(\xi_{tt}^{B}\right)^\mu\left(\xi_{tt}^{B}\right)_\mu = 2\pi^2\ell^2\Big[\cosh\!\left(2T_{\tilde U}\tilde U\right)\cosh\!\left(2T_{\tilde V}\tilde V\right) +\frac{\tilde\rho}{T_{\tilde U}T_{\tilde V}}\sinh\!\left(2T_{\tilde U}\tilde U\right)\sinh\!\left(2T_{\tilde V}\tilde V\right)-1\Big].
		\end{split}
	\end{equation}
	Between the horizons, the norm vanishes only on the curve \eqref{45}, and it is positive everywhere else. On the  outer horizon it is,
	\begin{equation}
		4\pi^2\ell^2\sinh^2\!\left(T_{\tilde U}\tilde U+T_{\tilde V}\tilde V\right)
	\end{equation}
	and on the inner horizon becomes
	\begin{equation}
		4\pi^2\ell^2\sinh^2\!\left(T_{\tilde U}\tilde U-T_{\tilde V}\tilde V\right).
	\end{equation}
	So the flow itself stays regular on the horizons, and only its components in the Rindler chart blow up.
	
\begin{figure}
		\centering
		\includegraphics[width= 0.4\textwidth]{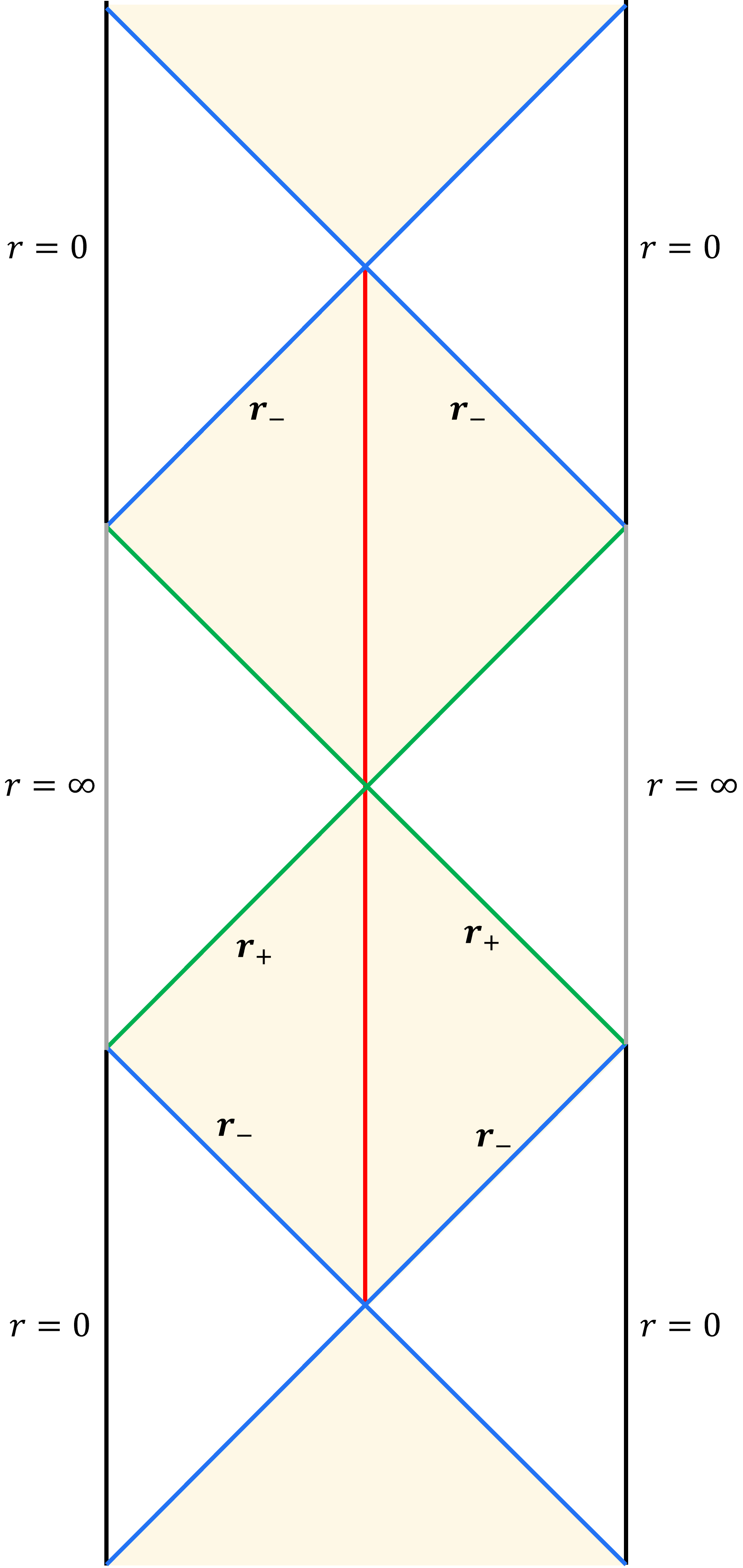}
		\caption{ \justifying Penrose diagram of the BTZ black hole. The green lines are the outer horizon $r_+$ and the blue lines are the inner horizon $r_-$. The red line is the timelike extremal curve \eqref{45}. It starts at the inner horizon, touches the outer horizon at its midpoint, and returns to the inner horizon.}
		\label{100}
	\end{figure}

	We rewrite the curve equations \eqref{30b} as
	\begin{equation}
		V=-\frac{l_V}{l_U}\,U, \qquad \rho=\frac{2 l_U}{l_V\left(l_U^2+4U^2\right)}.
	\end{equation}
	Substituting these into the inverse transformation for $\tilde\rho$, \eqref{46}, we get
	\begin{equation}
		\tilde\rho=T_{\tilde U}T_{\tilde V}\,\frac{l_U^2-4U^2}{l_U^2+4U^2}.
	\end{equation}
	In the above relation, $\tilde\rho$ goes from $-T_{\tilde U}T_{\tilde V}$ at $U\to-\infty$ to $T_{\tilde U}T_{\tilde V}$ at $U=0$, and back to $-T_{\tilde U}T_{\tilde V}$ at $U\to+\infty$. 
	Therefore, we find that the timelike curve starts at the inner horizon, touches the outer one at its midpoint $U=0$, and returns to the inner horizon, as shown in the Penrose diagram of BTZ in Fig.~\ref{100}.
	
	Under the map \eqref{35}, the line $\tilde U=\tilde V=0$ maps to the half $U\le0$ of the curve \eqref{30}. It is defined for every real $U$ and is symmetric under $U\to-U$, so the other half, $U\ge 0$, has the same length, and the whole curve has length
	\begin{equation}
		L_{tt}=2i\int_{-T_{\tilde U}T_{\tilde V}}^{T_{\tilde U}T_{\tilde V}}\frac{\ell\: d\tilde\rho}{2\:\sqrt{T_{\tilde U}^2T_{\tilde V}^2-\tilde\rho^2}}
		=i\pi\ell,
	\end{equation}
	where the factor $i$ reflects $ds^2<0$. This does not depend on $l_U$, $l_V$ or on the horizon parameters. With the Brown–Henneaux central charge $c=3\ell/\left(2G_N\right)$ we get
	\begin{equation}
		\frac{L_{tt}}{4G}=\frac{i\pi c}{6},
	\end{equation}
	which reproduces the imaginary part of the timelike entanglement entropy \cite{Doi2022,Doi_2023}.

	\section{Holographic Timelike Entanglement Entropy in Rindler Flat Space}
	
	In the previous section we gave the modular flows and their extremal curves a geometric meaning in Rindler BTZ. We now do the same in flat space and look for the object that each flow and curve corresponds to. The spacetime here is a flat-space cosmology (FSC), which has only a cosmological horizon. After the Rindler transformation, the spacelike extremal curve lies on this horizon. The timelike extremal curve, which is responsible for the imaginary part, stretches from the cosmological horizon to timelike infinity.  
	
	\subsection{TEE from Modular Flow in CCFT$_{2}$ on the Plane}
	\label{sec31}
	
\begin{figure}
		\centering
		\begin{minipage}{0.48\linewidth}
			\centering
			\includegraphics[width=\linewidth]{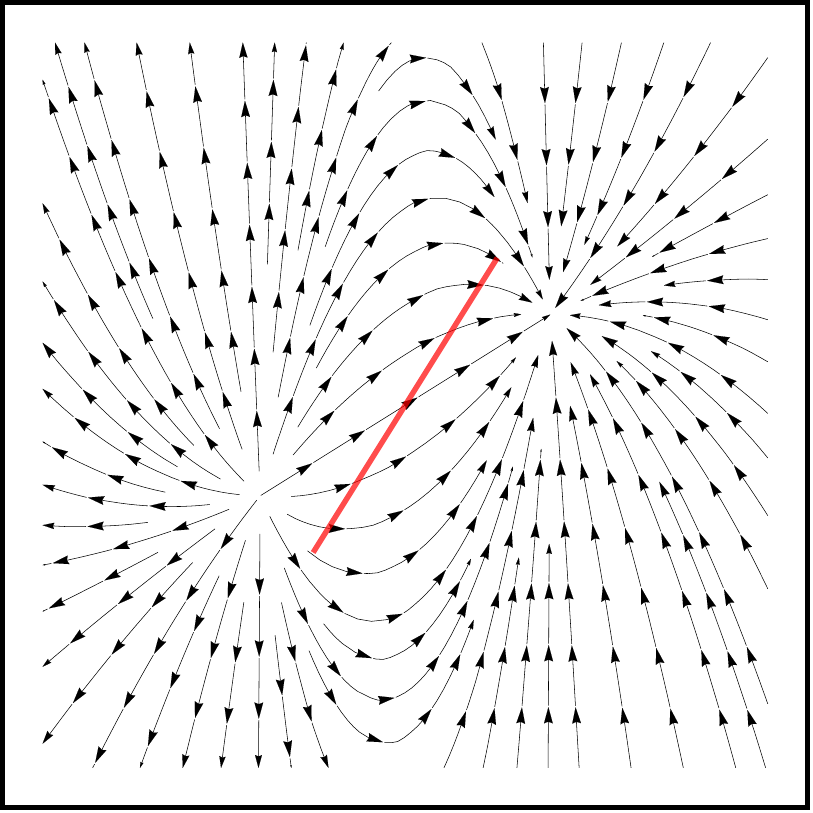}\\
			(a)
		\end{minipage}
		\hfill
		\begin{minipage}{0.48\linewidth}
			\centering
			\includegraphics[width=\linewidth]{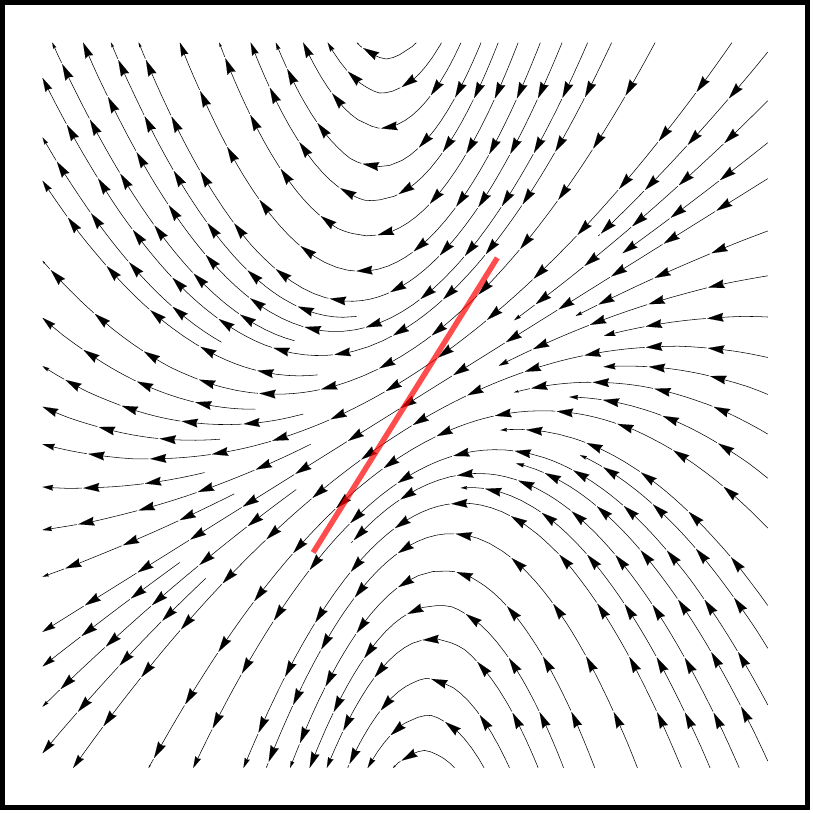}\\
			(b)
		\end{minipage}
		\caption{ \justifying Boundary modular flows of the CCFT$_2$ on the plane for (a) $\zeta^{b}_{st}$ and (b) $\zeta^{b}_{tt}$. In each panel the timelike interval is shown in red.}
		\label{101}
	\end{figure}

	To find the holographic TEE for a CCFT$_2$ on the plane, we build the modular flows from the symmetry generators of the theory. We work in Bondi coordinates, where the metric takes the form
	\begin{equation}
		ds^{2} = -2dudr + r^{2}d\phi^{2}.
		\label{16}
	\end{equation}
	The null boundary symmetry generators on the plane are given by 
	\begin{equation}
		\begin{aligned}
			L_{+1} &= -2u\phi\,\partial_u-\phi^2\,\partial_\phi,
			&\qquad
			M_{+1} &= \phi^2\,\partial_u,\\[0.5em]
			L_{0} &= -u\,\partial_u-\phi\,\partial_\phi,
			&\qquad
			M_{0} &= \phi\,\partial_u,\\[0.5em]
			L_{-1} &= -\partial_\phi,
			&\qquad
			M_{-1} &= \partial_u.
		\end{aligned}
	\end{equation}
	The corresponding bulk Killing vectors take the form
	\begin{equation}
		\begin{aligned}
			L_{+1} &= -2u\phi\,\partial_u + 2r\phi\,\partial_r + \left(\dfrac{2u}{r}-\phi^2\right)\partial_\phi, \\[0.5em]
			L_{0} &= -u\,\partial_u + r\,\partial_r - \phi\,\partial_\phi, \\[0.5em]
			L_{-1} &= -\partial_\phi, \\[1em]
			M_{+1} &= \phi^2\,\partial_u + 2\,\partial_r - \dfrac{2\phi}{r}\,\partial_\phi, \\[0.5em]
			M_{0} &= \phi\,\partial_u - \dfrac{1}{r}\,\partial_r, \\[0.5em]
			M_{-1} &= \partial_u.
		\end{aligned}
	\end{equation}
	With the boundary and bulk generators at hand, we turn to the modular flows of an interval on the null boundary characterized by the parameters $l_{u}$ and $l_{\phi}$. We introduce this interval as timelike because the flows constructed below coincide with the flat limit of the modular flows of the timelike interval in AdS$_3$.
	The boundary modular flows are then constructed as linear combinations of the boundary symmetry generators, and are shown in Fig.~\ref{101}. They are given by
	\begin{equation}
		\begin{split}
			\zeta_{st}^{b} ={}& \dfrac{\pi}{2 l_{\phi}} \left(4L_{+1}^{b} -l_{\phi}^{2} L_{-1}^{b} + l_{u} l_{\phi} M_{-1}^{b} + \dfrac{4 l_{u}}{l_{\phi}} M_{+1}^{b}\right) \\
			={}& \dfrac{\pi \left(l_{u} l_{\phi}^{2} - 8 l_{\phi} u \phi + 4 l_u \phi^{2}\right)}{2 l_{\phi}^{2}}\, \partial_{u} + \dfrac{\pi \left(l_{\phi}^{3} - 4 l_{\phi} \phi^{2}\right)}{2 l_{\phi}^{2}}\, \partial_{\phi}.
		\end{split}
	\end{equation}
	\begin{equation}
		\begin{split}
			\zeta_{tt}^{b} ={}& \dfrac{\pi}{2 l_{\phi}} \left(4L_{+1}^{b} + l_{\phi}^{2} L_{-1}^{b} - l_{u} l_{\phi} M_{-1}^{b} + \dfrac{4 l_{u}}{l_{\phi}} M_{+1}^{b}\right) \\
			={}& \dfrac{\pi \left(-l_{u} l_{\phi}^{2} - 8 l_{\phi} u \phi + 4 l_u \phi^{2}\right)}{2 l_{\phi}^{2}}\, \partial_{u} - \dfrac{\pi \left(l_{\phi}^{3} + 4 l_{\phi} \phi^{2}\right)}{2 l_{\phi}^{2}}\, \partial_{\phi}.
		\end{split}
	\end{equation}
	The corresponding bulk modular flows are obtained by replacing the boundary generators with their bulk counterparts, while keeping the coefficients unchanged. They are therefore 
	given by
	
	\begin{equation}
		\begin{split}
			\zeta_{st}^{B} ={}& \dfrac{\pi}{2 l_{\phi}} \left(4L_{+1}^{B} -l_{\phi}^{2} L_{-1}^{B} + l_{u} l_{\phi} M_{-1}^{B} + \dfrac{4 l_{u}}{l_{\phi}} M_{+1}^{B}\right) \\
			={}& \dfrac{\pi \left(l_{u} l_{\phi}^{2} - 8 l_{\phi} u \phi + 4 l_u \phi^{2}\right)}{2 l_{\phi}^{2}}\, \partial_{u} 
			+ \dfrac{4 \pi \left(l_{u} + l_{\phi} r \phi\right)}{l_{\phi}^{2}}\, \partial_{r} \\
			&+ \dfrac{\pi \left(l_{\phi}^{3} r + 8 l_{\phi} u - 8 l_{u} \phi - 4 l_{\phi} r \phi^{2}\right)}{2 l_{\phi}^{2} r}\, \partial_{\phi}.
		\end{split}
		\label{12}
	\end{equation}
	
	\begin{equation}
		\begin{split}
			\zeta_{tt}^{B} ={}& \dfrac{\pi}{2 l_{\phi}} \left(4L_{+1}^{B} +l_{\phi}^{2} L_{-1}^{B} - l_{u} l_{\phi} M_{-1}^{B} + \dfrac{4 l_{u}}{l_{\phi}} M_{+1}^{B}\right) \\
			={}& \dfrac{\pi \left(-l_{u} l_{\phi}^{2} - 8 l_{\phi} u \phi + 4 l_u \phi^{2}\right)}{2 l_{\phi}^{2}}\, \partial_{u} 
			+ \dfrac{4 \pi \left(l_{u} + l_{\phi} r \phi\right)}{l_{\phi}^{2}}\, \partial_{r} \\
			&+ \dfrac{\pi \left(-l_{\phi}^{3} r + 8 l_{\phi} u - 8 l_{u} \phi - 4 l_{\phi} r \phi^{2}\right)}{2 l_{\phi}^{2} r}\, \partial_{\phi}.
		\end{split}
		\label{13}
	\end{equation}
	
	It should be noted that the flat space modular flows constructed above from the boundary and bulk symmetry generators of CCFT$_2$ and flat$_{3}$ can also be obtained as the flat space limit of the AdS$_3$/CFT$_2$ modular flows derived in \eqref{6}, \eqref{7}, \eqref{11} and \eqref{49}.
	
	We now determine the extremal curves associated with the bulk modular flows, since these give the holographic entanglement entropy. Setting the components of $\zeta^B_{st}$ in \eqref{12} to zero, all three vanish on the spacelike extremal curve
	\begin{equation}
		\begin{split}
			\phi = {}& -\dfrac{l_{u}}{r l_{\phi}}, \\[0.5ex]
			u ={}& - \dfrac{r l_{\phi}^{2}}{8} - \dfrac{l_{u}^{2}}{2 r l_{\phi}^{2}}.
			\label{14}
		\end{split}
	\end{equation}
	
	With the flat space metric given in \eqref{16} and the spacelike extremal curve obtained in \eqref{14} the line element along the spacelike curve is readily found to be 
	\begin{equation}
		ds_{\zeta_{st}}^{f} = \dfrac{l_{\phi}}{2} dr.
	\end{equation}
	The line element alone does not fix the length; we also need the integration range. Unlike in AdS, where the spacelike extremal curve is anchored at the endpoints of the interval, the flat space extremal curve does not reach the boundary. It is instead joined to the endpoints by two null rays extending to null infinity \cite{Jiang_2017}.
	Each ray meets the curve at one point, and only the segment between these points contributes to the entropy, since $ds^2=0$ along the rays. The two branches join at $r=0$, which we take as the lower limit of integration. The upper limit $r=r_I^f$ (with $f$ for flat and $I$ for intersection) is the radius at which the curve meets the null ray through the endpoint of the interval. It is fixed by $du/dr=0$, which gives
	\begin{equation}
		r_{I}^f = \dfrac{2 l_{u}}{l_{\phi}^{2}}. 
	\end{equation}
	Integrating the line element, we obtain
	\begin{equation}
		L_{st} = 2 \int_{0}^{r_{I}^f} ds_{\zeta_{st}}^{f} =  \dfrac{2 l_{u}}{l_{\phi}}.
	\end{equation}
	The prefactor 2 arises because the two branches of the curve give equal contributions.
	
	Similarly, the timelike extremal curve corresponding to the bulk modular flow in \eqref{13} is
	\begin{equation}
		\begin{split}
			\phi = {}&-\dfrac{l_{u}}{r l_{\phi}}, \\ 
			u ={}&  \dfrac{r l_{\phi}^{2}}{8} - \dfrac{l_{u}^{2}}{2 r l_{\phi}^{2}}.
			\label{15}
		\end{split}
	\end{equation}
	Following a similar procedure and using the timelike curve equation \eqref{15} the corresponding line element is given by
	\begin{equation}
		ds_{\zeta_{tt}}^{f} = i \dfrac{l_{\phi}}{2} dr.
	\end{equation}
	Integrating over the same range gives
	\begin{equation}
		L_{tt} = 2 \int_{0}^{r_{I}^f} ds_{\zeta_{tt}}^{f} = i \dfrac{2  l_{u}}{l_{\phi}}.
	\end{equation}
	Adding the two contributions, the total length associated with the timelike interval on the plane is
	\begin{equation}
		L_{tot} = L_{st} + L_{tt} =  \left(1 + i\right) \dfrac{2 l_{u}}{l_{\phi}}.
		\label{22}
	\end{equation}
	Consequently, the holographic timelike entanglement entropy is
	\begin{equation}
		S^{(T)}=\dfrac{L_{tot}}{4G}= \dfrac{l_u}{2 G l_\phi}+i \dfrac{l_u}{2 G l_\phi}.
			\label{66}
	\end{equation}
	Notably, the real and imaginary parts of the TEE for a timelike interval in CCFT$_2$ on the plane are equal. In contrast, this does not hold for a timelike interval on the cylinder \cite{shalamzari2026modularflowholographictimelike}. For small $l_\phi$ the cylinder result of \cite{shalamzari2026modularflowholographictimelike} reduces to \eqref{66}.
	
	We now examine whether the flat space entropy can be recovered as a limit of the corresponding AdS result. One might expect that taking $\ell\to\infty$ in \eqref{17} would reproduce the flat space expression. However, this quantity does not admit a well-defined limit. To circumvent this problem, we follow the prescription of \cite{Fareghbal:2024lqa}, in which the entropy is computed from the segment of the extremal curve between $r=0$ and $r=r_I$, rather than from the entire curve. To implement it, we first determine $r_I$ in AdS$_3$ in BMS coordinates. Eliminating $\phi$ between the two equations of \eqref{24}, we can express $u$ in terms of $r$ as
	\begin{equation}
		u=\dfrac{\ell \left(2 \ell r \pm \dfrac{l_\phi \sqrt{r^2 \left(l_u^2 - \ell^2 l_\phi^2\right) \left(r^2 l_u^2 - r^2 \ell^2 l_\phi^2 -4 \ell^4\right)}}{\ell^2 l_\phi^2 -l_u^2}\right)}{2 r^2}.
		\label{27}
	\end{equation}
	We denote by $r_I$ the radius at which $u$ attains its extremum along the spacelike extremal curve. By extremizing \eqref{27} with respect to $r$, we find
	\begin{equation}
		r_{I} = \dfrac{2 \ell^{2} l_{u}}{\ell^{2} l_{\phi}^{2} - l_{u}^{2}}.
		\label{53}
	\end{equation}
	Integrating the line element given in \eqref{18} and \eqref{19} yields
	\begin{equation}
		\begin{split}
			L = 2 \int_{0}^{r_{I}} ds_{\xi_{st}} + 2 \int_{0}^{r_{I}} ds_{\xi_{tt}} 
			= 2 \ell \sinh^{-1} \left(\dfrac{l_{u}}{\sqrt{\ell^{2} l_{\phi}^{2} - l_u^{2}}}\right) + 2 i \ell \sin^{-1} \left(\dfrac{l_{u}}{\sqrt{\ell^{2} l_{\phi}^{2} - l_u^{2}}}\right).
		\end{split}
		\label{20}
	\end{equation}
	Taking the flat limit of  \eqref{20}, we recover the real and imaginary parts of \eqref{22} and therefore the corresponding TEE in the context of flat space holography.
	
	\subsection{Holographic Timelike Entanglement Entropy in Rindler FSC}
	\label{sec32}
	
	We have found the geometric meaning of modular flows and their extremal curves for a timelike interval in the Rindler BTZ. Next, we pose the same question in flat space. Which geometric object does each flow and curve correspond to? In particular, what is the meaning of the timelike curve, which is responsible for the imaginary part? The spacetime we consider is a flat-space cosmology (FSC), which possesses only a cosmological horizon. We start with the Rindler transformations in flat space \cite{Jiang_2017}, which are given by
	\begin{widetext}
		\begin{equation}
			\begin{split}
				\tilde{r} ={}& \sqrt{\dfrac{\tilde{M}}{16 l_\phi^2}\left(8u-4l_u+r(l_\phi-2\phi)^2\right)\left(8u+4l_u+r(l_\phi+2\phi)^2\right)+\dfrac{\tilde{J}^2}{4\tilde{M}}}, \\[2ex]
				\tilde{\phi} ={}& -\dfrac{1}{\sqrt{\tilde{M}}}\log\left[\dfrac{\sqrt{\tilde{M}}}{4 l_\phi}\,\dfrac{8u-4l_u+r(l_\phi-2\phi)^2}{\tilde{r}+\tilde{J}/(2\sqrt{\tilde{M}})}\right], \\[2ex]
				\tilde{u} ={}& \dfrac{\tilde{r}}{\tilde{M}}+\dfrac{1}{4 l_\phi \sqrt{\tilde{M}}}\left(8u+4r\phi^2-r l_\phi^2\right)-\dfrac{\tilde{J}}{2\tilde{M}}\,\tilde{\phi}.
			\end{split}
			\label{52}
		\end{equation}
	\end{widetext}
	
	Inverting these, we find
	
	\begin{widetext}
		\begin{equation}
			\begin{split}
				r ={}& \dfrac{1}{2 l_\phi \tilde{M}}\left[\left(2\tilde{r}\sqrt{\tilde{M}}-\tilde{J}\right)e^{\sqrt{\tilde{M}}\tilde{\phi}}+\left(2\tilde{r}\sqrt{\tilde{M}}+\tilde{J}\right)e^{-\sqrt{\tilde{M}}\tilde{\phi}}-2\sqrt{\tilde{M}}\left(2\tilde{M}\tilde{u}-2\tilde{r}+\tilde{J}\tilde{\phi}\right)\right], \\[2ex]
				\phi ={}& \dfrac{1}{r}\left[\dfrac{\left(2\tilde{r}\sqrt{\tilde{M}}-\tilde{J}\right)e^{\sqrt{\tilde{M}}\tilde{\phi}}-\left(2\tilde{r}\sqrt{\tilde{M}}+\tilde{J}\right)e^{-\sqrt{\tilde{M}}\tilde{\phi}}}{4\tilde{M}}-\dfrac{l_u}{l_\phi}\right], \\[2ex]
				u ={}& \dfrac{1}{8}\left[\left(l_\phi^2-4\phi^2\right)r+\dfrac{2 l_\phi}{\sqrt{\tilde{M}}}\left(2\tilde{M}\tilde{u}-2\tilde{r}+\tilde{J}\tilde{\phi}\right)\right].
			\end{split}
		\end{equation}
	\end{widetext}
	These transformations map the flat metric \eqref{16} to an FSC in Bondi coordinates with the metric
	\begin{equation}
		ds^2=\tilde{M}\,d\tilde{u}^2-2\,d\tilde{u}\,d\tilde{r}+\tilde{J}\,d\tilde{u}\,d\tilde{\phi}+\tilde{r}^2\,d\tilde{\phi}^2 ,
		\label{50}
	\end{equation}
	where $\tilde{M}$ and $\tilde{J}$ are the mass and angular momentum parameters of the FSC. Its cosmological horizon lies at $\tilde{r}_c=|\tilde{J}|/(2\sqrt{\tilde{M}})$.
	
	Applying these transformations to the spacelike and timelike modular flows $\zeta_{st}$ and $\zeta_{tt}$ in \eqref{12} and \eqref{13}, we obtain, in the tilde coordinates,
	
	\begin{equation}
		\hspace{-9.7cm}
		\widetilde{\zeta_{st}^{B}}= -\dfrac{\pi \tilde{J}}{\tilde{M}^{3/2}}\, \partial_{\tilde{u}} + \dfrac{2\pi}{\sqrt{\tilde{M}}}\, \partial_{\tilde{\phi}}.
	\end{equation}
	\begin{widetext}
		\begin{equation}
			\begin{split}
				\widetilde{\zeta_{tt}^{B}}= {} & \left(\dfrac{\pi \tilde{J}}{\tilde{M}^{3/2}}\cosh\!\left(\sqrt{\tilde{M}}\,\tilde{\phi}\right)-\dfrac{\pi}{\tilde{M}}\sinh\!\left(\sqrt{\tilde{M}}\,\tilde{\phi}\right)\left(2\tilde{M}\tilde{u}+\tilde{J}\tilde{\phi}\right)\right)\partial_{\tilde{u}} \\[0.9ex]
				& +\dfrac{\pi\left(\tilde{J}\cosh\!\left(\sqrt{\tilde{M}}\,\tilde{\phi}\right)-2\sqrt{\tilde{M}}\sinh\!\left(\sqrt{\tilde{M}}\,\tilde{\phi}\right)\tilde{r}\right)\left(2\tilde{M}\tilde{u}-2\tilde{r}+\tilde{J}\tilde{\phi}\right)}{2\sqrt{\tilde{M}}\,\tilde{r}}\,\partial_{\tilde{r}} \\[0.9ex]
				& +\dfrac{\pi\cosh\!\left(\sqrt{\tilde{M}}\,\tilde{\phi}\right)\left(2\tilde{M}\tilde{u}-2\tilde{r}+\tilde{J}\tilde{\phi}\right)}{\sqrt{\tilde{M}}\,\tilde{r}}\,\partial_{\tilde{\phi}}.
			\end{split}
		\end{equation}
	\end{widetext}
	
	The flow $\widetilde{\zeta_{st}^{B}}$ has constant components and therefore never vanishes. We instead examine its norm, which can vanish where the flow becomes null, as happens on a Killing horizon. The norm is
	\begin{equation}
		\left(\widetilde{\zeta_{st}^{B}}\right)^\mu \left(\widetilde{\zeta_{st}^{B}}\right)_\mu=\frac{4\pi^2}{\tilde M}\left(\tilde r^2-\tilde r_c^2\right),
	\end{equation}
	which vanishes exactly at the cosmological horizon $\tilde r=\tilde r_c$. Hence $\widetilde{\zeta_{st}^{B}}$ is a Killing vector whose Killing horizon is the cosmological horizon; it is spacelike for $\tilde r>\tilde r_c$ and timelike for $\tilde r<\tilde r_c$. The extremal curve of $\widetilde{\zeta_{st}^{B}}$, the locus where the flow becomes null, therefore lies on the cosmological horizon.
	
\begin{figure}
		\centering
		
		\begin{subfigure}[t]{0.4\linewidth}
			\centering
			\includegraphics[width=\linewidth]{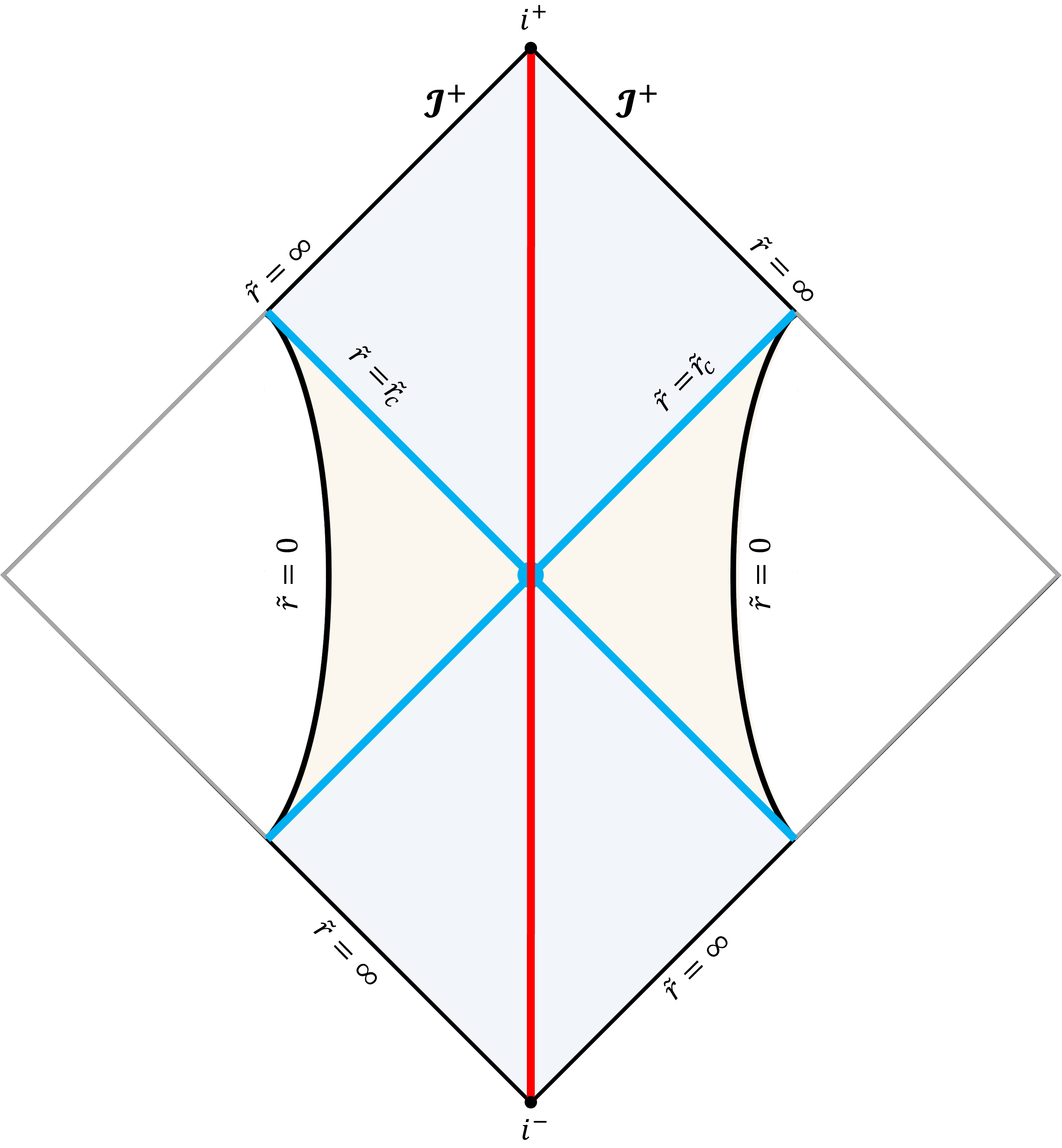}
			\caption{ \justifying The timelike curve stretched from $i^-$ to $i^+$, and is shown in red. The Rindler horizon, which the spacelike curve lies on, is shown in blue.}
			\label{106}
		\end{subfigure}
		\hfill
		\begin{subfigure}[t]{0.4\linewidth}
			\centering
			\includegraphics[width=\linewidth]{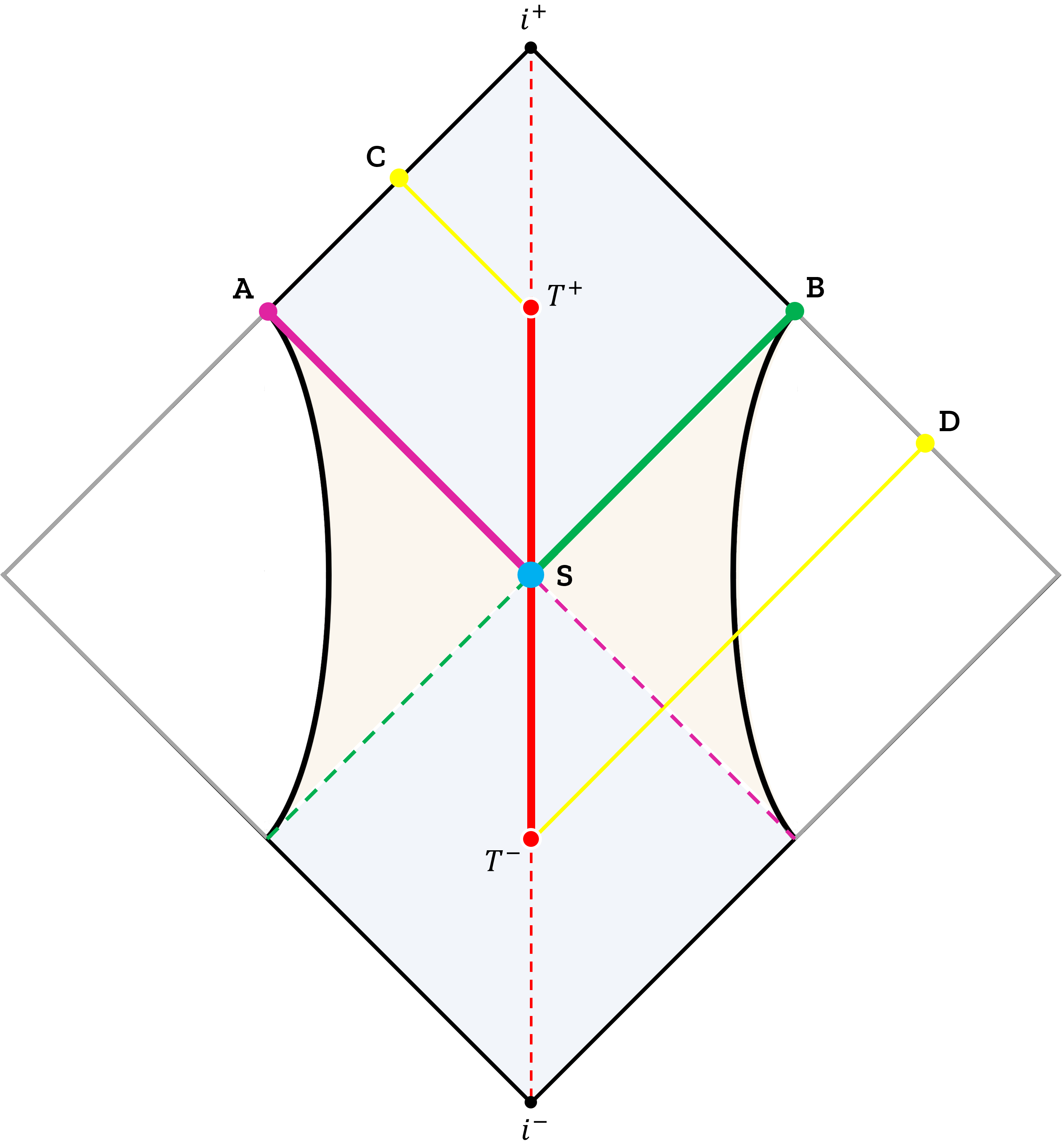}
			\caption{ \justifying Only the stretch from $T^-$ to $T^+$ through the bifurcation point $S$ enters $L_{tt}$. The spacelike curve lies on the bifurcation line. Its null rays stay on the horizon and reach null infinity at $A$ and $B$, whereas the rays from $T^\pm$ leave the horizon and end at $C$ and $D$.}
			\label{107}
		\end{subfigure}
		
		\vspace{0.4cm}
		\begin{subfigure}[t]{0.6\linewidth}
			\centering
			\includegraphics[width=\linewidth]{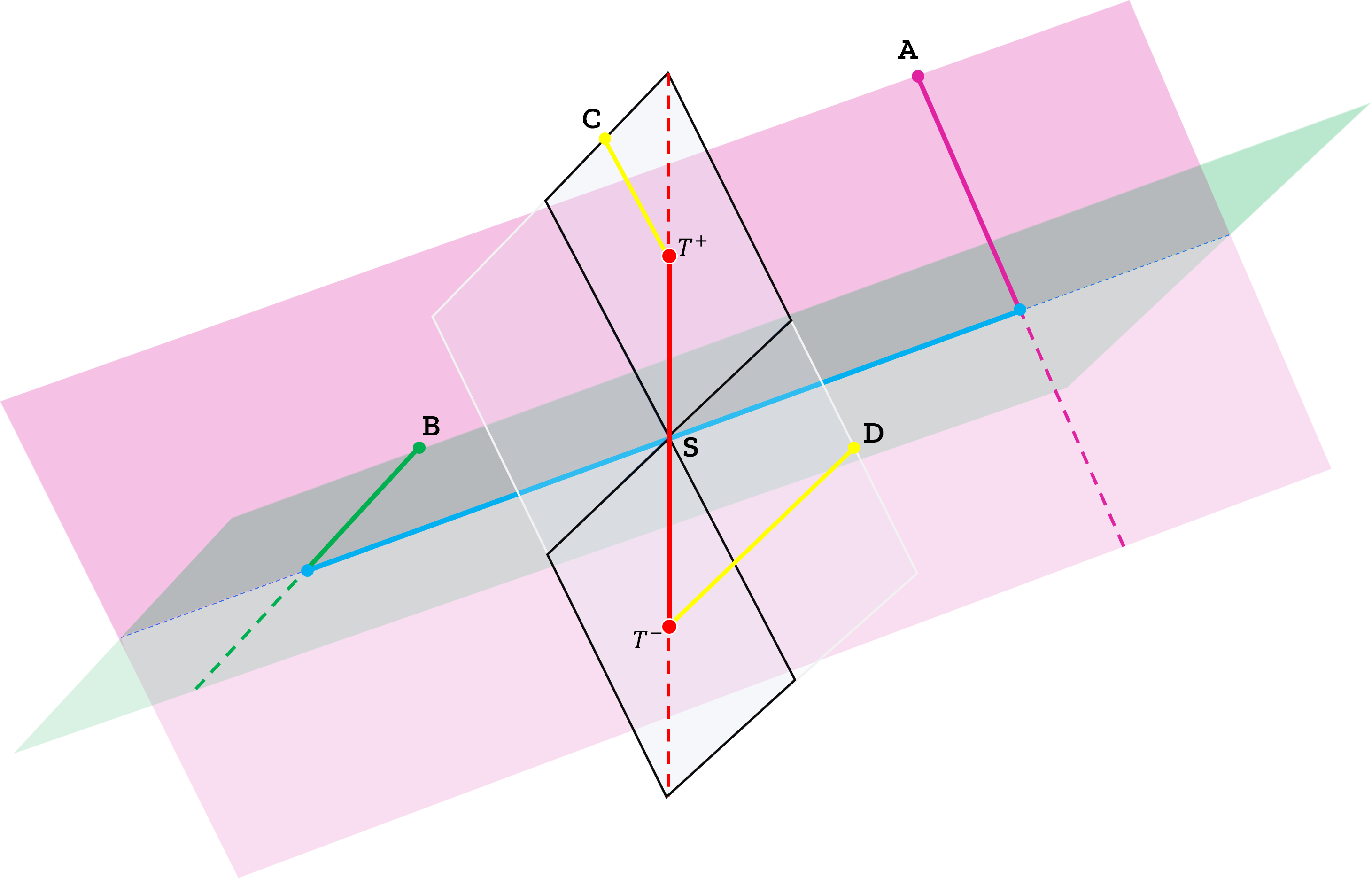}
			\caption{ \justifying The two null sheets of the horizon, unfolded.}
			\label{108}
		\end{subfigure}
		
		\caption{ \justifying Extremal curves of the timelike entanglement entropy in the Penrose diagram of the flat space cosmology}
		\label{105}
	\end{figure}

	Since $\widetilde{\zeta_{tt}^{B}}$ has points where it vanishes, its extremal curve is obtained by setting its components to zero:
	\begin{equation}
		\begin{split}
			\tilde{\phi}= {}&\frac{1}{2\sqrt{\tilde{M}}}\,\ln\left(\frac{2\sqrt{\tilde{M}}\tilde{r}+\tilde{J}}{2\sqrt{\tilde{M}}\tilde{r}-\tilde{J}}\right), \\
			\tilde{u}= {}&\frac{\tilde{r}}{\tilde{M}}-\frac{\tilde{J}}{4\tilde{M}^{3/2}}\ln\,\left(\frac{2\sqrt{\tilde{M}}\tilde{r}+\tilde{J}}{2\sqrt{\tilde{M}}\tilde{r}-\tilde{J}}\right).
			\label{51}
		\end{split}
	\end{equation}
	Both are real only for $\tilde r>\tilde r_c=|\tilde J|/(2\sqrt{\tilde M})$. As $\tilde r\to\tilde r_c$ the curve approaches the cosmological horizon asymptotically, and as $\tilde r\to\infty$ it runs out to infinity. This agrees with the BTZ case. In the flat limit the inner horizon of BTZ becomes the cosmological horizon, while the outer horizon moves to infinity. In section~\ref{sec22} we showed that the timelike curve of $\widetilde{\xi_{tt}^{B}}$ runs between the two horizons. It is therefore natural that in flat space the timelike curve of $\widetilde{\zeta_{tt}^{B}}$ runs between the cosmological horizon and timelike infinity, as demonstrated in the Penrose diagram of FSC in Fig.~\ref{105}.
	
	In Fig.~\ref{105}(a) the cosmological horizon is depicted in blue. The spacelike extremal curve
	lies within this bifurcate horizon. Notably, the timelike extremal curve stretches between the cosmological horizon and timelike infinity and is shown in red. In panel~\ref{105}(b) the timelike curve is cut down to the part that enters $L_{tt}$, the solid red stretch from $T^-$ to $T^+$ through the bifurcation point $S$, of total proper length $2l_u/l_\phi$. The faint red dashed ends still reach $i^\pm$ but contribute nothing. The null rays leaving $T^\pm$ go elsewhere, to $C$ and $D$ at $(0,\mp l_\phi/2)$ in the same order. The one from $T^+$ stays inside the expanding region; the one from $T^-$ crosses the horizon and meets $\tilde r=0$ on the way, so $D$ falls in the hatched part that exists only in the covering space. Changing $\tilde J$ moves $\tilde r=0$ and that crossing, while every labelled point keeps its place.
	
	The spacelike curve \eqref{14} lies on the bifurcation line, which this section squeezes into $S$. Null rays from its endpoints run along the horizon $\tilde r=\tilde r_c$ and end at $A$ and $B$, the images of the interval endpoints $(u,\phi)=(\mp l_u/2,\mp l_\phi/2)$ on future null infinity, in that order. Finally, in Fig.~\ref{105}(c) the two null sheets of the horizon are unfolded.
	
	Substituting the equations of the timelike curve \eqref{51} into the metric \eqref{50}, we find the line element
	\begin{equation}
		ds^2=\frac{4\tilde{r}^2}{\tilde{J}^2-4\tilde{M}\tilde{r}^2}\,d\tilde{r}^2 .
	\end{equation}
	The line element depends only on $\tilde r$, so we need the range of $\tilde r$ along the curve. As shown above, the curve starts at the horizon of the flat-space cosmology (FSC), which gives the lower limit $\tilde r_c=|\tilde J|/(2\sqrt{\tilde M})$. The upper limit is
	\begin{equation}
		\tilde r_I^f=\sqrt{\frac{\tilde M\,l_u^2}{l_\phi^2}+\frac{\tilde J^2}{4\tilde M}},
		\label{55}
	\end{equation}
	the image under the first Rindler transformation \eqref{52} of the point $(u,r,\phi)=\left(0,\,2l_u/l_\phi^2,\,-l_\phi/2\right)$. The integral therefore runs from $\tilde r_c$ to $\tilde r_I^f$. The factor of 2 accounts for the two branches of the curve, which contribute equally:
	
	\begin{equation}
		\widetilde{L}_{tt} = 2i\int_{\tilde{r}_c}^{\tilde{r}_I^f} \frac{2\tilde{r}}{\sqrt{4\tilde{M}\tilde{r}^2-\tilde{J}^2}}\,d\tilde{r}
		= \frac{2i\,l_u}{l_\phi}.
	\end{equation}	
	This length reproduces the imaginary part of TEE for timelike interval in CCFT$_2$ on the plane.
	
	As a consistency check, we map the point $I$ from AdS$_3$ to the Rindler coordinates and ask whether the flat-limit result \eqref{55} comes back. We start from $r_I$ in \eqref{53}. Using \eqref{29} and \eqref{54}, we find $\rho_I$ to be
	\begin{equation}
		\rho_I = \dfrac{\left(l_U - l_V\right)^2}{2\, l_U^2\, l_V^2}.
	\end{equation}
	Now we set $\tilde U=\tilde V=0$ in the first relation of the Rindler transformations \eqref{35} and insert $\rho_I$. Solving for $\tilde\rho$ gives
	\begin{equation}
		\tilde{\rho}_I = \dfrac{\left(l_U^2 - 4\, l_U l_V + l_V^2\right) T_{\tilde{U}} T_{\tilde{V}}}{2\, l_U l_V}.
	\end{equation}
	This is not yet the radial coordinate. In the Bondi coordinates, which suit the flat limit, $\tilde r$ is related to $\tilde\rho$ by \cite{Jiang_2017}
	\begin{equation}
		\tilde{r} = \dfrac{\sqrt{2\tilde{\rho} + T_{\tilde{V}}^2 + T_{\tilde{U}}^2}}{2}.
	\end{equation}
	The horizons enter through $T_{\tilde U}$ and $T_{\tilde V}$. In terms of $\tilde M$ and $\tilde J$, they are
	\begin{equation}
		\begin{split}
			\tilde{r}_{+} ={}& \sqrt{\frac{\ell}{2}\left(\tilde{M}\ell + \sqrt{\tilde{M}^2\ell^2 - \tilde{J}^2}\right)}, \\[0.9ex]
			\tilde{r}_{-} ={}& \sqrt{\frac{\ell}{2}\left(\tilde{M}\ell - \sqrt{\tilde{M}^2\ell^2 - \tilde{J}^2}\right)}.
		\end{split}
	\end{equation}
	Using \eqref{56} together with these horizon radii, we get
	\begin{equation}
		\tilde{r}_I = \sqrt{\dfrac{\tilde{J}^2}{2\left(\tilde{M} + \sqrt{\tilde{M}^2 - \dfrac{\tilde{J}^2}{\ell^2}}\right)} + \dfrac{l_u^2 \sqrt{\tilde{M}^2 - \dfrac{\tilde{J}^2}{\ell^2}}}{l_\phi^2 - \dfrac{l_u^2}{\ell^2}}}.
	\end{equation}
	In the flat limit $\ell\to\infty$, the expression reduces to $\tilde r_I^f$ given by \eqref{55}. The point $I$, carried from AdS$_3$ through the Rindler coordinates to the BMS gauge, therefore lands at the same place as in the direct flat-space calculation and thus the two routes agree.

	\section{Discussion}
	\label{sec5}
	
	 Our aim was to locate the imaginary part of the holographic timelike entanglement entropy in the Rindler picture, and to describe the timelike extremal curve in a geometry where the spacelike curves lie on horizons. We have done this for a timelike interval, in Rindler BTZ and in a flat-space cosmology.	We studied holographic timelike entanglement entropy (TEE) through the modular flow of a timelike interval, first in AdS$_3$ and then in flat space. In both cases the extremal curves are the fixed points of the bulk modular flows, and a Rindler transformation shows where these curves sit.
	
	For a CFT$_2$ on the plane, we wrote the boundary and bulk flows in Poincar\'e and BMS coordinates. The flow $\xi_{st}$ vanishes on a spacelike curve with two branches, and $\xi_{tt}$ vanishes on a timelike curve that joins them. Their lengths give the real and the imaginary part of the entropy, and the sum agrees with the known result. The third flow, $\xi_{ss}$, belongs to a spacelike interval and does not enter the entropy. It still helps with the geometry, because it is exchanged with $\xi_{st}$ under $l_x\leftrightarrow l_t$.

	The Rindler part is where the geometry shows up. Under the transformation to a rotating BTZ black hole, $\xi_{st}$ and $\xi_{ss}$ become constant Killing vectors. Their norms vanish exactly on the inner and the outer horizon respectively, so the spacelike extremal curves lie on the horizons, and the exchange $l_u\leftrightarrow \ell l_\phi$ turns out to be a passage between the inner and the outer horizons. The timelike curve is of a different kind. It is the line $\tilde U=\tilde V=0$ across the region between the horizons, a geodesic because it is the fixed-point set of the isometry generated by $\xi_{tt}$. It starts on the inner horizon, touches the outer one at its midpoint and returns to the inner horizon. Its length is $i\pi\ell$ for every interval and every choice of horizon radii, so the imaginary part $i\pi c/6$ is fixed by the black hole geometry between the horizons and by nothing else. This is the piece that a thermal entropy cannot provide, and it is the gap left in \cite{He2024,Wen2025}. Our $\xi_{tt}$ complements \cite{Wen2025}, where the pre-image of the inner horizon was identified with the spacelike geodesics of the TEE, while the timelike geodesic was found not to be a fixed-point set of the modular momentum flow. Here it is the zero set of $\xi_{tt}$. 
	
	For a CCFT$_2$ on the plane, we built the flows $\zeta_{st}$ and $\zeta_{tt}$ from the BMS symmetry generators. The spacelike and timelike curves have lengths $2l_u/l_\phi$ and $i\,2l_u/l_\phi$, so the real and imaginary parts of the entropy are equal. This does not hold on the cylinder \cite{shalamzari2026modularflowholographictimelike}. For small $l_\phi$ the cylinder result $l_u\cot(l_\phi/2)+i\,l_u\csc(l_\phi/2)$ of \cite{shalamzari2026modularflowholographictimelike} reduces to the plane result, since both $\cot(l_\phi/2)$ and $\csc(l_\phi/2)$ approach $2/l_\phi$. Taking $\ell\to\infty$ in the AdS entropy gives no finite limit when we use the whole curve. We followed \cite{Fareghbal:2024lqa} and kept only the part between $r=0$ and $r_I$, and then the flat limit gives the flat space result.
	
	In Rindler flat space we used a flat-space cosmology (FSC) with a single cosmological horizon. The spacelike curve lies on this horizon, and the timelike curve runs from the horizon to timelike infinity. This matches the BTZ picture. In the flat limit the inner horizon becomes the cosmological horizon, and the outer horizon goes to infinity. The length of the timelike curve is $i\,2l_u/l_\phi$, which reproduces the imaginary part of the entropy. As a check, we carried the point $I$ from AdS$_3$ through the Rindler coordinates. In the flat limit it lands at the same $\tilde r_I$ that we found directly in flat space.
	
	Several questions remain open. Our argument identifies the curve and its length, but it does not identify the bulk dual of the transition matrix whose pseudo entropy this is, a problem that stays open in the pseudo entropy interpretation \cite{Nakata2021,Doi_2023,Murciano_2022}; a proposal based on the modular momentum is made in \cite{Wen2025}. The construction should also be repeated with the corresponding modular flows on the cylinder, where the real and imaginary parts no longer coincide in flat space, and at finite temperature. Higher dimensions would show whether the timelike curve still sits between a pair of horizons, and the same question can be asked for states dual to a rotating BTZ background, where the horizons are physical and not only a Rindler frame. Finally, how all this relates to the black hole interior probes of \cite{LiYang2026}, where a Cauchy horizon controls the transition between timelike and spacelike entanglement, is a natural next question.

	\appendix
	\renewcommand{\theequation}{\Alph{section}\arabic{equation}}
	\section{Absence of Timelike extremal curve beyond the horizons}
	\label{appendix}
	
	We investigate whether the modular flow admits zeros at any point outside the region between the two horizons. To this end, the Rindler transformation corresponding to each of the two regions is considered separately, for the region outside the outer horizon, $\tilde\rho>\tilde\rho_+ =T_{\tilde U}T_{\tilde V}$, and for the region inside the inner horizon, $\tilde\rho<\tilde\rho_- = -T_{\tilde U}T_{\tilde V}$. These regions are denoted by $>\tilde\rho_+$ and $<\tilde\rho_-$, respectively.
	
	Inside the inner horizon, $\tilde\rho<\tilde\rho_-$, the  Rindler transformation is given by
	
	\begin{widetext}
		\begin{equation}
			\begin{split}
				\rho ={}& \left(\sqrt{\dfrac{-\left(T_{\tilde{U}} T_{\tilde{V}} + \tilde{\rho}\right)}{l_U l_V T_{\tilde{U}} T_{\tilde{V}} }} \sinh\left(T_{\tilde{U}} \tilde{U} + T_{\tilde{V}} \tilde{V}\right) - \sqrt{\dfrac{T_{\tilde{U}} T_{\tilde{V}} - \tilde{\rho}}{l_U l_V T_{\tilde{U}} T_{\tilde{V}} }} \sinh\left(T_{\tilde{U}} \tilde{U} - T_{\tilde{V}} \tilde{V}\right) \right)^2, \\[3ex]
				U ={}& \dfrac{\sqrt{\dfrac{-l_U \left(T_{\tilde{U}} T_{\tilde{V}} + \tilde{\rho}\right)}{4 l_V T_{\tilde{U}} T_{\tilde{V}}}} \cosh\left(T_{\tilde{U}} \tilde{U} + T_{\tilde{V}} \tilde{V}\right)  - \sqrt{\dfrac{l_U \left(T_{\tilde{U}} T_{\tilde{V}} - \tilde{\rho}\right)}{4 l_V T_{\tilde{U}} T_{\tilde{V}}}} \cosh\left(T_{\tilde{U}} \tilde{U} - T_{\tilde{V}} \tilde{V}\right)}{\sqrt{\dfrac{-\left(T_{\tilde{U}} T_{\tilde{V}} + \tilde{\rho}\right)}{l_U l_V T_{\tilde{U}} T_{\tilde{V}} }} \sinh\left(T_{\tilde{U}} \tilde{U} + T_{\tilde{V}} \tilde{V}\right) - \sqrt{\dfrac{T_{\tilde{U}} T_{\tilde{V}} - \tilde{\rho}}{l_U l_V T_{\tilde{U}} T_{\tilde{V}} }} \sinh\left(T_{\tilde{U}} \tilde{U} - T_{\tilde{V}} \tilde{V}\right) }, \\[3ex]
				V ={}& \dfrac{\sqrt{\dfrac{-l_V \left(T_{\tilde{U}} T_{\tilde{V}} + \tilde{\rho}\right)}{4 l_U T_{\tilde{U}} T_{\tilde{V}}}} \cosh\left(T_{\tilde{U}} \tilde{U} + T_{\tilde{V}} \tilde{V}\right)  + \sqrt{\dfrac{l_V \left(T_{\tilde{U}} T_{\tilde{V}} - \tilde{\rho}\right)}{4 l_U T_{\tilde{U}} T_{\tilde{V}}}} \cosh \left(T_{\tilde{U}} \tilde{U} - T_{\tilde{V}} \tilde{V}\right)}{\sqrt{\dfrac{-\left(T_{\tilde{U}} T_{\tilde{V}} + \tilde{\rho}\right)}{l_U l_V T_{\tilde{U}} T_{\tilde{V}} }} \sinh\left(T_{\tilde{U}} \tilde{U} + T_{\tilde{V}} \tilde{V}\right) - \sqrt{\dfrac{T_{\tilde{U}} T_{\tilde{V}} - \tilde{\rho}}{l_U l_V T_{\tilde{U}} T_{\tilde{V}} }} \sinh\left(T_{\tilde{U}} \tilde{U} - T_{\tilde{V}} \tilde{V}\right) }.
			\end{split}
		\end{equation}
	\end{widetext}
	The inverse transformations are
	\begin{widetext}
		\begin{equation}
			\begin{split}
				\tilde{U} ={}& \dfrac{\log \left[\dfrac{\left(2+ \left(l_U + 2 U\right) \rho \left(l_V + 2V\right)\right) \left(2- \left(l_U + 2U\right) \rho \left(l_V - 2V\right)\right)}{\left(2+ \left(l_U-2U\right) \rho \left(l_V-2V\right)\right) \left(2- \left(l_U - 2U\right) \rho \left(l_V + 2V\right)\right)}\right]}{4 T_{\tilde{U}}}, \\[3ex]
				\tilde{V} ={}& \dfrac{\log \left[\dfrac{\left(2+ \left(l_U+2U\right) \rho \left(l_V+2V\right)\right) \left(2- \left(l_U - 2U\right) \rho \left(l_V + 2V\right)\right)}{\left(2+ \left(l_U - 2U\right) \rho \left(l_V - 2V\right)\right) \left(2- \left(l_U+2U\right) \rho \left(l_V-2V\right)\right)}\right]}{4 T_{\tilde{V}}}, \\[3ex]
				\tilde{\rho} ={}& \dfrac{T_{\tilde{U}} T_{\tilde{V}} \left(4 +16 U V \rho  + \left(l_U^2 - 4 U^2\right) \left(l_V^2 - 4 V^2\right) \rho^2\right)}{4 l_U l_V \rho}.
			\end{split}
		\end{equation}
	\end{widetext}
	Applying these transformations to the modular flow $\xi^{B}_{tt}$ in \eqref{43}, we obtain
	\begin{widetext}
		\begin{equation}
			\begin{split}
				\xi_{<\tilde{\rho}_-}^{B}= {} & -4\pi \sqrt{\tilde{\rho}^2-T_{\tilde{U}}^2 T_{\tilde{V}}^2}\, \sinh\!\left(T_{\tilde{U}} \tilde{U}-T_{\tilde{V}} \tilde{V}\right) \cosh\!\left(T_{\tilde{U}} \tilde{U}+T_{\tilde{V}} \tilde{V}\right) \partial_{\tilde{\rho}} \\
				& +\dfrac{\pi\left(\tilde{\rho}\, \cosh\!\left(2 T_{\tilde{U}} \tilde{U}\right)+T_{\tilde{U}} T_{\tilde{V}} \cosh\!\left(2 T_{\tilde{V}} \tilde{V}\right)\right)}{T_{\tilde{U}} \sqrt{\tilde{\rho}^2-T_{\tilde{U}}^2 T_{\tilde{V}}^2}}\, \partial_{\tilde{U}} \\
				& -\dfrac{\pi\left(T_{\tilde{U}} T_{\tilde{V}} \cosh\!\left(2 T_{\tilde{U}} \tilde{U}\right)+\tilde{\rho}\, \cosh\!\left(2 T_{\tilde{V}} \tilde{V}\right)\right)}{T_{\tilde{V}} \sqrt{\tilde{\rho}^2-T_{\tilde{U}}^2 T_{\tilde{V}}^2}}\, \partial_{\tilde{V}}.
			\end{split}
		\end{equation}
	\end{widetext}
	Setting the components of $\xi^{B}_{<\tilde\rho_-}$ to zero leads to
	
	\begin{equation}
		T_{\tilde{U}} \tilde{U}=T_{\tilde{V}} \tilde{V}, \qquad	\cosh\!\left(2 T_{\tilde{U}} \tilde{U}\right) \left(\tilde{\rho}+T_{\tilde{U}} T_{\tilde{V}} \right)=0.
	\end{equation}
	The second equation cannot be satisfied, so the flow has no zeros in this region.
	
	For the outer horizon, in the region $\tilde\rho>\tilde\rho_+$, the Rindler transformation reads
	
	\begin{widetext}
		\begin{equation}
			\begin{split}
				\rho ={}& \left(\sqrt{\dfrac{\tilde{\rho} + T_{\tilde{U}} T_{\tilde{V}}}{l_U l_V T_{\tilde{U}} T_{\tilde{V}} }} \cosh\left(T_{\tilde{U}} \tilde{U} + T_{\tilde{V}} \tilde{V}\right) + \sqrt{\dfrac{\tilde{\rho} - T_{\tilde{U}} T_{\tilde{V}}}{l_U l_V T_{\tilde{U}} T_{\tilde{V}} }} \cosh\left(T_{\tilde{U}} \tilde{U} - T_{\tilde{V}} \tilde{V}\right) \right)^2, \\[3ex]
				U ={}& \dfrac{\sqrt{\dfrac{l_U \left(\tilde{\rho} + T_{\tilde{U}} T_{\tilde{V}}\right)}{4 l_V T_{\tilde{U}} T_{\tilde{V}}}} \sinh\left(T_{\tilde{U}} \tilde{U} + T_{\tilde{V}} \tilde{V}\right)  + \sqrt{\dfrac{l_U \left(\tilde{\rho} - T_{\tilde{U}} T_{\tilde{V}}\right)}{4 l_V T_{\tilde{U}} T_{\tilde{V}}}} \sinh\left(T_{\tilde{U}} \tilde{U} - T_{\tilde{V}} \tilde{V}\right)}{\sqrt{\dfrac{\tilde{\rho} + T_{\tilde{U}} T_{\tilde{V}}}{l_U l_V T_{\tilde{U}} T_{\tilde{V}} }} \cosh\left(T_{\tilde{U}} \tilde{U} + T_{\tilde{V}} \tilde{V}\right) + \sqrt{\dfrac{\tilde{\rho} - T_{\tilde{U}} T_{\tilde{V}}}{l_U l_V T_{\tilde{U}} T_{\tilde{V}} }} \cosh\left(T_{\tilde{U}} \tilde{U} - T_{\tilde{V}} \tilde{V}\right) }, \\[3ex]
				V ={}& \dfrac{\sqrt{\dfrac{l_V \left(\tilde{\rho} + T_{\tilde{U}} T_{\tilde{V}}\right)}{4 l_U T_{\tilde{U}} T_{\tilde{V}}}} \sinh\left(T_{\tilde{U}} \tilde{U} + T_{\tilde{V}} \tilde{V}\right)  + \sqrt{\dfrac{l_V \left(\tilde{\rho} - T_{\tilde{U}} T_{\tilde{V}}\right)}{4 l_U T_{\tilde{U}} T_{\tilde{V}}}} \sinh \left(T_{\tilde{V}} \tilde{V} - T_{\tilde{U}} \tilde{U}\right)}{\sqrt{\dfrac{\tilde{\rho} + T_{\tilde{U}} T_{\tilde{V}}}{l_U l_V T_{\tilde{U}} T_{\tilde{V}} }} \cosh\left(T_{\tilde{U}} \tilde{U} + T_{\tilde{V}} \tilde{V}\right) + \sqrt{\dfrac{\tilde{\rho} - T_{\tilde{U}} T_{\tilde{V}}}{l_U l_V T_{\tilde{U}} T_{\tilde{V}} }} \cosh\left(T_{\tilde{U}} \tilde{U} - T_{\tilde{V}} \tilde{V}\right) }.
			\end{split}
		\end{equation}
	\end{widetext}
	The inverse transformations, taken from \cite{Jiang_2017}, are
	\begin{widetext}
		\begin{equation}
			\begin{split}
				\tilde{U} ={}& \dfrac{\log \left[\dfrac{\left(2+ \left(l_U + 2 U\right) \rho \left(l_V + 2V\right)\right) \left(2- \left(l_U + 2U\right) \rho \left(l_V - 2V\right)\right)}{\left(2+ \left(l_U-2U\right) \rho \left(l_V-2V\right)\right) \left(2- \left(l_U - 2U\right) \rho \left(l_V + 2V\right)\right)}\right]}{4 T_{\tilde{U}}}, \\[3ex]
				\tilde{V} ={}& \dfrac{\log \left[\dfrac{\left(2+ \left(l_U+2U\right) \rho \left(l_V+2V\right)\right) \left(2- \left(l_U - 2U\right) \rho \left(l_V + 2V\right)\right)}{\left(2+ \left(l_U - 2U\right) \rho \left(l_V - 2V\right)\right) \left(2- \left(l_U+2U\right) \rho \left(l_V-2V\right)\right)}\right]}{4 T_{\tilde{V}}}, \\[3ex]
				\tilde{\rho} ={}& \dfrac{T_{\tilde{U}} T_{\tilde{V}} \left(4 +16 U V \rho  + \left(l_U^2 - 4 U^2\right) \left(l_V^2 - 4 V^2\right) \rho^2\right)}{4 l_U l_V \rho}.
			\end{split}
		\end{equation}
	\end{widetext}
	
	The modular flow $\xi^{B}_{tt}$ in \eqref{43} becomes
	\begin{widetext}
		\begin{equation}
			\begin{split}
				\xi_{>\tilde{\rho}_{+}}^{B}= {} & 4\pi \sqrt{\tilde{\rho}^2-T_{\tilde{U}}^2 T_{\tilde{V}}^2}\, \sinh\!\left(T_{\tilde{U}} \tilde{U}+T_{\tilde{V}} \tilde{V}\right) \cosh\!\left(T_{\tilde{U}} \tilde{U}-T_{\tilde{V}} \tilde{V}\right) \partial_{\tilde{\rho}} \\[1.7ex]
				& -\dfrac{\pi\left(\tilde{\rho}\, \cosh\!\left(2 T_{\tilde{U}} \tilde{U}\right)-T_{\tilde{U}} T_{\tilde{V}} \cosh\!\left(2 T_{\tilde{V}} \tilde{V}\right)\right)}{T_{\tilde{U}} \sqrt{\tilde{\rho}^2-T_{\tilde{U}}^2 T_{\tilde{V}}^2}}\, \partial_{\tilde{U}} \\[1.5ex]
				& +\dfrac{\pi\left(T_{\tilde{U}} T_{\tilde{V}} \cosh\!\left(2 T_{\tilde{U}} \tilde{U}\right)-\tilde{\rho}\, \cosh\!\left(2 T_{\tilde{V}} \tilde{V}\right)\right)}{T_{\tilde{V}} \sqrt{\tilde{\rho}^2-T_{\tilde{U}}^2 T_{\tilde{V}}^2}}\, \partial_{\tilde{V}}.
			\end{split}
		\end{equation}
	\end{widetext}
	The equations of the curve associated with the modular flow $\xi^{B}_{>\tilde\rho_+}$ are
	\begin{equation}
		T_{\tilde{U}} \tilde{U}=-T_{\tilde{V}} \tilde{V}, \qquad	\cosh\!\left(2 T_{\tilde{U}} \tilde{U}\right) \left(\tilde{\rho}-T_{\tilde{U}} T_{\tilde{V}} \right)=0.
	\end{equation}
	These relations show that $\xi^{B}_{>\tilde\rho_+}$ has no fixed points for $\tilde\rho>\tilde\rho_+$.
	We have thus shown that neither inside the inner horizon nor outside the outer horizon is there a curve on which the modular flow $\xi^{B}_{tt}$ in \eqref{43} vanishes after the Rindler transformation. The timelike extremal curve therefore exists only in the region between the two horizons, as we found in \eqref{45}.
	
			\begin{acknowledgments}
				
			We are grateful to Reza Fareghbal for helpful remarks on the first draft of this paper. MHSH was partly supported by Shahid Chamran University of Ahvaz under grant no.\ SCU.SP1404.86612. MKZ thanks Shahid Chamran University of Ahvaz, Iran, for supporting this work under research grant no.\ SCU.SP1404.37271.
		
	\end{acknowledgments}

	\nocite{REVTEX42Control,apsrev42Control}
	\bibliographystyle{apsrev4-2}
	\bibliography{refrences2}
	
\end{document}